\documentclass[english,onecolumn,draftcls]{IEEEtran}
\usepackage[T1]{fontenc}
\usepackage[latin9]{inputenc}
\usepackage{geometry}
\usepackage{amsthm}
\usepackage{amsmath}
\usepackage{amssymb}
\usepackage{graphicx}
\usepackage{esint}

\makeatletter
\theoremstyle{plain}
\newtheorem{thm}{\protect\theoremname}
  \theoremstyle{plain}
  \newtheorem{cor}{\protect\corollaryname}
  \theoremstyle{plain}
  \newtheorem{lem}{\protect\lemmaname}

\usepackage{cite}
\usepackage{enumerate}
\usepackage{amsmath}
\usepackage{flushend}

\newcommand{\openone}{\leavevmode\hbox{\small1\normalsize\kern-.33em1}}

\makeatother

\usepackage{babel}
  \providecommand{\lemmaname}{Lemma}
\providecommand{\corollaryname}{Corollary}
\providecommand{\theoremname}{Theorem}

\begin{document}

\title{On the Dispersions of the Gel'fand-Pinsker Channel and Dirty Paper
Coding}

\author{Jonathan Scarlett}

\maketitle
\long\def\symbolfootnote[#1]#2{\begingroup\def\thefootnote{\fnsymbol{footnote}}\footnote[#1]{#2}\endgroup}
\begin{abstract}
This paper studies second-order coding rates for memoryless channels
with a state sequence known non-causally at the encoder. In the case
of finite alphabets, an achievability result is obtained using constant-composition
random coding, and by using a small fraction of the block to transmit
the type of the state sequence. For error probabilities less than
$\frac{1}{2}$, it is shown that the second-order rate improves on
an existing one based on i.i.d. random coding. In the Gaussian case
(dirty paper coding) with an almost-sure power constraint, an achievability
result is obtained used using random coding over the surface of a
sphere, and using a small fraction of the block to transmit a quantized
description of the state power. It is shown that the second-order
asymptotics are identical to the single-user Gaussian channel of the
same input power without a state. \end{abstract}
\begin{IEEEkeywords}
Channels with state, Gel'fand-Pinsker channel, dirty paper coding,
channel dispersion, second-order coding rate
\end{IEEEkeywords}
\symbolfootnote[0]{J. Scarlett is with the Department of Engineering, University of Cambridge, Cambridge, CB2 1PZ, U.K. (e-mail: jmscarlett@gmail.com).}

\section{Introduction \label{sec:INTRO}}

The problem of characterizing the second-order asymptotics of the
highest achievable channel coding rate at a given error probability
and increasing block length was studied by Strassen \cite{Strassen},
and has recently regained significant attention following the works
of Polyanskiy \emph{et al. }\cite{Finite} and Hayashi \cite{Hayashi}.
For discrete memoryless channels, the maximum number of codewords
$M^{*}(n,\epsilon)$ of length $n$ yielding an average error probability
not exceeding $\epsilon$ satisfies
\begin{equation}
\log M^{*}(n,\epsilon)=nC-\sqrt{nV}\mathsf{Q}^{-1}(\epsilon)+o(\sqrt{n}),\label{eq:GP_SingleUser0}
\end{equation}
where $C$ is the channel capacity, $\mathsf{Q}^{-1}(\cdot)$ is the
inverse of the $\mathsf{Q}$-function, and $V$ is known as the \emph{channel
dispersion}. We can interpret $C$ and $V$ as being the mean and
variance of the information density $i(x,y)\triangleq\log\frac{W(y|x)}{\sum_{\overline{x}}Q(\overline{x})W(y|\overline{x})}$
for some capacity-achieving input distribution $Q$. For the additive
white Gaussian noise (AWGN) channel with maximal power $P$, an expansion
of the form \eqref{eq:GP_SingleUser0} holds with $V=\frac{P(2+P)}{2(1+P)^{2}}$
\cite{Finite,Hayashi}.

\begin{figure}
\begin{centering}
\includegraphics[width=0.5\paperwidth]{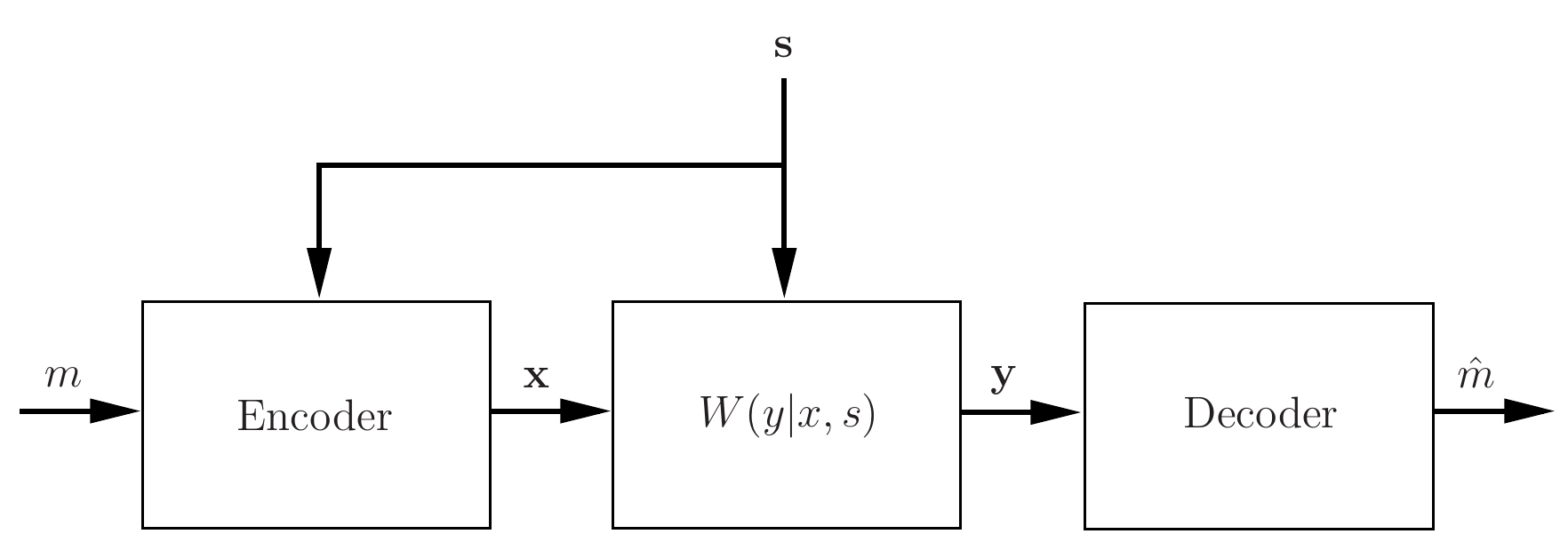}
\par\end{centering}

\caption{Channel model.}

\end{figure}

In this paper, we study the second-order asymptotics of channel coding
with a random state known non-causally at the encoder, as studied
by Gel'fand-Pinsker \cite{GelfPinsk} and Costa \cite{DirtyPaper}
(see Figure 1). In the case of finite alphabets and unconstrained
inputs, we give an achievability result of the form \eqref{eq:GP_SingleUser0},
with $\ge$ in place of the equality. In the case that the channel
is Gaussian and the input is subject to an almost-sure power constraint
(dirty paper coding \cite{DirtyPaper}), we show that the second-order
asymptotics are identical to those obtained when the state is absent,
thus strengthening the well-known analogous result for the capacity.

\subsection{Channel Model and Capacity \label{sub:GP_CHANNEL}}

The alphabets of the input, output and state are denoted by $\mathcal{X}$,
$\mathcal{Y}$ and $\mathcal{S}$ respectively. The channel is assumed
to be memoryless with a transition law $W(y|x,s)$, and the state
sequence $\boldsymbol{S}=(S_{1},\cdots,S_{n})$ is distributed according
to $P_{\boldsymbol{S}}(\boldsymbol{s})$. The $n$-letter channel
transition law is given by 
\begin{equation}
W^{n}(\boldsymbol{y}|\boldsymbol{x},\boldsymbol{s})\triangleq\prod_{i=1}^{n}W(y_{i}|x_{i},s_{i}).
\end{equation}
The encoder takes as input the state sequence $\boldsymbol{s}$ and
a message $m$ equiprobable on the set $\{1,\cdots,M\}$, and transmits
a codeword $\boldsymbol{x}^{(m)}(\boldsymbol{s})$. The decoder forms
an estimate $\hat{m}$ based on $\boldsymbol{y}$, and an error is
said to have occurred if $\hat{m}\ne m$. We study two variations
of this setup, which we refer to as the discrete case and the Gaussian
case.

In the discrete case, the alphabets $\mathcal{X}$, $\mathcal{Y}$
and $\mathcal{S}$ are assumed to be finite, the channel input is
assumed to be unconstrained, and the state distribution is assumed
to be i.i.d. on some distribution $\pi$, i.e. $P_{\boldsymbol{S}}(\boldsymbol{s})=\prod_{i=1}^{n}\pi(s_{i})$.
The capacity is given by \cite{GelfPinsk} 
\begin{equation}
C=\max_{\mathcal{U},Q_{U|S},\phi(\cdot,\cdot)}I(U;Y)-I(U;S),\label{eq:GP_Capacity}
\end{equation}
where the mutual informations are computed using the distribution
\begin{equation}
P_{SUY}(s,u,y)=\pi(s)Q_{U|S}(u|s)W(y|\phi(u,s),s)\label{eq:GP_Distr}
\end{equation}
and the maximum is over all finite alphabets $\mathcal{U}$, conditional
distributions $Q_{U|S}$ and functions $\phi\,:\,\mathcal{U}\times\mathcal{S}\to\mathcal{X}$.

In the Gaussian case, the channel is described by
\begin{equation}
\boldsymbol{Y}=\boldsymbol{X}+\boldsymbol{S}+\boldsymbol{Z},\label{eq:DPC_Channel}
\end{equation}
where $\boldsymbol{Z}$ is an i.i.d. noise sequence with $Z_{i}\sim N(0,1)$.
That is, we have
\begin{align}
W(y|x,s) & =\frac{1}{\sqrt{2\pi}}e^{-\frac{(y-x-s)^{2}}{2}}.\label{eq:DPC_Channel2}
\end{align}
The state distribution $P_{\boldsymbol{S}}$ is assumed to be arbitrary
for now. The codewords are constrained to satisfy a power constraint
of the form
\begin{equation}
\|\boldsymbol{x}^{(m)}(\boldsymbol{s})\|^{2}\le nP,\qquad\forall m,\boldsymbol{s}\label{eq:DPC_InputConstr}
\end{equation}
for some transmit power $P$. That is, we require $\|\boldsymbol{X}\|^{2}\le nP$
almost surely. This is a stricter constraint than that considered
in some previous works; see Section \ref{sub:DPC_CONSTRAINTS} for
further discussion.

Here we provide an outline of the capacity results; see \cite{DirtyPaper},
\cite[Sec. 7.7]{NetworkBook} for details. We first consider the case
that $P_{\boldsymbol{S}}$ is i.i.d. on $\pi\sim N(0,P_{\pi})$ for
some $P_{\pi}>0$. The capacity is given by \eqref{eq:GP_Capacity}
subject to the constraint $\mathbb{E}[\phi(U,S)^{2}]\le P$. We fix
$\alpha>0$ and choose
\begin{align}
Q_{U|S}(\cdot|s) & \sim N(-\alpha s,P)\label{eq:DPC_OptQ}\\
\phi(u,s) & =u-\alpha s,\label{eq:DPC_OptPhi}
\end{align}
which can be equivalently be written as 
\begin{align}
U & =X+\alpha S\label{eq:DPC_U}\\
X & \sim N(0,P),\label{eq:DPC_X}
\end{align}
where $X$ is independent of $S$. Under these parameters, it can
be shown that
\begin{align}
I(U;Y) & =\frac{1}{2}\log\Bigg(\frac{(P+\Pi+1)(P+\alpha^{2}\Pi)}{P\Pi(1-\alpha)^{2}+(P+\alpha^{2}\Pi)}\Bigg)\label{eq:DPC_IUY}\\
I(U;S) & =\frac{1}{2}\log\Bigg(\frac{P+\alpha^{2}\Pi}{P}\Bigg).\label{eq:DPC_IUS}
\end{align}
Furthermore, the optimal choice of $\alpha$ is given by
\begin{equation}
\alpha=\frac{P}{1+P},\label{eq:DPC_alpha}
\end{equation}
and yields
\begin{equation}
C=\frac{1}{2}\log(1+P).\label{eq:DPC_C}
\end{equation}
Thus, the capacity is independent of the state power $\Pi$, and is
the same as if the state sequence were absent (or equivalently, if
it were known at the decoder). 

In the case of a non-Gaussian i.i.d. state sequence with $\mathbb{E}[S_{i}^{2}]<\infty$,
the capacity remains the same. For example, see \cite[Sec. 7.7]{NetworkBook}
for a proof based on connections with minimum mean square error (MMSE)
estimation. Although we consider a general (possibly non-Gaussian
and non-i.i.d.) state distribution in this paper, the parameter choices
and mutual informations in \eqref{eq:DPC_OptQ}--\eqref{eq:DPC_alpha}
will play a major role in the analysis.

\subsection{Discussion: Power Constraints and Common Randomness for Dirty Paper
Coding \label{sub:DPC_CONSTRAINTS}}

In general, the fundamental performance limits of channels with power
constraints can vary depending on (i) the type of power constraint
(e.g. almost-sure vs. averaged over a random variable), (ii) the availability
of common randomness at the encoder and decoder, and (iii) whether
the average or maximal error probability is being considered. Here
we focus on the case of average error probability, and discuss some
variations of dirty paper coding in which the former two properties
differ. For each case we consider, the capacity will remain equal
to \eqref{eq:DPC_C}, at least subject to mild technical conditions
on $P_{\boldsymbol{S}}$.

Suppose that the power constraint is averaged over the randomness
of the message and the state. In this case, we can show that the strong
converse fails to hold, similarly to the AWGN channel without state
\cite[Sec. 4.3.3]{FiniteThesis}. Since the capacity is given by \eqref{eq:DPC_C},
there exists a code of average power $\frac{P}{1-\epsilon}$, rate
approaching $\frac{1}{2}\log\big(1+\frac{P}{1-\epsilon}\big)$, and
vanishing error probability. By replacing the fraction $\epsilon$
of the codewords $\boldsymbol{x}^{(m)}(\boldsymbol{s})$ with the
highest power (averaged over $\boldsymbol{S}$) by the all-zero codeword,
we obtain a code of average power not exceeding $P$, rate approaching
$\frac{1}{2}\log\big(1+\frac{P}{1-\epsilon}\big)$, and error probability
approaching $\epsilon$. Since the strong converse does hold under
a maximal power constraint \cite{FiniteThesis}, we conclude that
the existence of a code satisfying an average power constraint does
not, in general, imply the existence of a code satisfying a maximal
power constraint and having the same asymptotic rate and error probability.

The study of lattice coding for the dirty paper coding problem generally
makes use of common randomness at the encoder and decoder in the form
of a dither; see \cite{LatticeCapacity,ExpLattice,DispLattice} and
references therein. The power constraint considered in these works
holds for all messages and state sequences, but it is averaged over
the randomness of the dither. To the best of the author's judgment,
the removal of this common randomness (if possible) would require
relaxing the power constraint to be averaged over the message and
state, and would thus recover the setting discussed in the previous
paragraph in which the strong converse fails to hold.

As seen in \eqref{eq:DPC_InputConstr}, the setting we consider is
stricter in the sense that the power constraint is an almost-sure
constraint, and no common randomness is assumed. This is the same
setup as that considered in \cite{DirtyPaper,ArbitraryDirty}, among
others.

\subsection{Previous Work}

For unconstrained channels with state known at the encoder, Watanabe
\emph{et al.} \cite{TanSideInfo} and Yassaee \emph{et al}. \cite{OneShot}
provided alternative derivations of the same result using different
techniques based on i.i.d. random coding. In order to state the result,
we introduce some definitions. We say that a triplet $(n,M,\epsilon)$
is achievable if there exists a code with block length $n$ containing
at least $M$ messages and yielding an average error probability not
exceeding $\epsilon$, and we define
\begin{equation}
M^{*}(n,\epsilon)\triangleq\max\Big\{ M\,:\,(n,M,\epsilon)\text{ is achievable}\Big\}.
\end{equation}
Letting $P_{Y|U}$, $P_{Y}$, etc. denote the marginals of \eqref{eq:GP_Distr},
we define the information densities
\begin{align}
i(u,s) & \triangleq\log\frac{Q_{U|S}(u|s)}{P_{U}(u)}\label{eq:GP_ius}\\
i(u,y) & \triangleq\log\frac{P_{Y|U}(y|u)}{P_{Y}(y)}\label{eq:GP_iuy}
\end{align}
with a slight abuse of notation. Furthermore, for a $2\times2$ positive
semi-definite matrix $\boldsymbol{V}$, we define the set
\begin{equation}
\mathsf{Q}_{\mathrm{inv}}(\boldsymbol{V},\epsilon)\triangleq\Big\{\boldsymbol{z}\in\mathbb{R}^{2}\,:\,\mathbb{P}\big[\boldsymbol{Z}\preceq\boldsymbol{z}]\ge1-\epsilon\Big\},\label{eq:GP_Qinv2}
\end{equation}
where $\boldsymbol{Z}\sim N(\boldsymbol{0},\boldsymbol{V})$, and
$\preceq$ denotes element-wise inequality. It was shown in \cite{TanSideInfo,OneShot}
that 
\begin{equation}
\log M^{*}(n,\epsilon)\ge nC-\sqrt{n}\tilde{R}+O(\log n),\label{eq:GP_Existing}
\end{equation}
where
\begin{align}
\tilde{R} & \triangleq\min_{(\tilde{R}_{1},\tilde{R}_{2})\in\mathsf{Q}_{\mathrm{inv}}(\boldsymbol{V},\epsilon)}\tilde{R}_{1}+\tilde{R}_{2}\\
\boldsymbol{V} & \triangleq\mathrm{Cov}\left[\begin{array}{c}
-i(U,S)\\
i(U,Y)
\end{array}\right].
\end{align}
For the case that an input constraint is present (e.g. dirty paper
coding), a similar expansion was provided in \cite{TanSideInfo} using
a $3\times3$ covariance matrix $\boldsymbol{V}$, with the third
entry added to capture the probability that the random i.i.d. codeword
violates the constraint.

A study of the second-order asymptotics of the modulo-lattice additive
noise channel was provided by Jiang and Liu \cite{DispLattice}. By
a data-processing argument, their result provides an achievable second-order
expansion of the rate for dirty paper coding with common randomness
at the encoder and decoder, and with a power constraint which is averaged
over the common randomness. In particular, \cite[Thm. 1]{DispLattice}
bears a strong resemblance to Theorem \ref{thm:DPC_MainResult} below.
However, it should be noted that our setting assumes a stricter power
constraint without common randomness, as discussed in Section \ref{sub:DPC_CONSTRAINTS}.
Furthermore, the analysis in \cite{DispLattice} is significantly
different from ours.

For related work on random-coding error exponents, see \cite{GPExponent1,GPExponent2,ExpLattice}
and references therein.

\subsection{Contributions}

As stated previously, the main contributions of this paper are a second-order
achievability result for the discrete case, and a conclusive characterization
of the second-order asymptotics for the Gaussian case. In the discrete
case with a target error probability less than $\frac{1}{2}$, we
show that our result can be weakened to \eqref{eq:GP_Existing}. For
the Gaussian case, we show that the dispersion is the same as that
of the AWGN channel of the same input power without a state.

Our result for the discrete case is based on constant-composition
random coding, which has recently been shown to yield gains in the
second-order performance of other network information theory problems
\cite{PaperMAC_2ord,OutputStatsRB}. In the Gaussian case, we use
a variant of random coding according to a uniform distribution on
a shell, which has been used for the single-user Gaussian channel
\cite{Finite} and the Gaussian multiple-access channel \cite{MolavianJazi}.
In both cases, we reduce the problem to that of a genie-aided setting
by using a small fraction of the block length to inform the decoder
of a property of the state sequence, namely, its empirical distribution
or its quantized power (e.g. see \cite{GPExponent1}). A key part
of our analysis for the discrete case makes use of techniques recently
introduced by Tomamichel and Tan \cite[Lemmas 17-18]{TanState}.

\subsection{\label{sub:MD_NOTATION}Notation}

Bold symbols are used for vectors and matrices (e.g. $\boldsymbol{x}$),
and the corresponding $i$-th entry of a vector is denoted with a
subscript (e.g. $x_{i}$). Given two vectors, say $\boldsymbol{x}_{1}$
and $\boldsymbol{x}_{2}$, we define the inner product $\langle\boldsymbol{x}_{1},\boldsymbol{x}_{2}\rangle=\sum_{i}x_{1,i}x_{2,i}$,
the $\ell_{2}$-norm $\|\boldsymbol{x}_{1}\|=\sqrt{\langle\boldsymbol{x}_{1},\boldsymbol{x}_{1}\rangle}$,
and the $\ell_{\infty}$-norm $\|\boldsymbol{x}_{1}\|_{\infty}=\max_{i}|x_{1,i}|$. 

The marginals of a joint distribution $P_{XY}$ are denoted by $P_{X}$
and $P_{Y}$. Probability, expectation and variance are respectively
denoted by $\mathbb{P}[\cdot]$, $\mathbb{E}[\cdot]$, and $\mathrm{Var}[\cdot]$.
When the meaning is clear, we will use shorthands such as $\mathbb{P}[\cdot\,|\, x]$
to denote conditioning on events such as $X=x$. For two sequences
$f_{n}$ and $g_{n}$, we write $f_{n}=O(g_{n})$ if $|f_{n}|\le c|g_{n}|$
for some $c$ and sufficiently large $n$, $f_{n}=o(g_{n})$ if $\lim_{n\to\infty}\frac{f_{n}}{g_{n}}=0$,
and $f_{n}=\Theta(g_{n})$ if both $f_{n}=O(g_{n})$ and $g_{n}=O(f_{n})$
hold.

\section{Statement of Main Results }

In this section, we present a formal statement of our main results,
along with some discussions and comparisons to existing results. 
\begin{thm}
\label{thm:GP_MainResult} Consider a discrete-memoryless Gel'fand-Pinsker
channel described by $\pi(s)$ and $W(y|x,s)$. Let $\mathcal{U}$,
$Q_{U|S}$ and $\phi(u,s)$ by any set of capacity-achieving parameters
in \eqref{eq:GP_Capacity}, and let $P_{SUY}$, $i(u,s)$ and $i(u,y)$
be as given in \eqref{eq:GP_Distr}, \eqref{eq:GP_ius} and \eqref{eq:GP_iuy}
under these parameters. If $\mathbb{E}\big[\mathrm{Var}[i(U,Y)\,|\, S,U]\big]>0$,
then 
\begin{equation}
\log M^{*}(n,\epsilon)\ge nC-\sqrt{nV}\mathsf{Q}^{-1}(\epsilon)+O(\log n),\label{eq:GP_MainResult}
\end{equation}
where
\begin{align}
V & \triangleq\mathbb{E}\big[\mathrm{Var}[i(U,Y)\,|\, S,U]\big]+\mathrm{Var}\big[\mathbb{E}[i(U,Y)-i(U,S)\,|\, S]\big]\label{eq:GP_V}\\
 & =\mathrm{Var}\big[i(U,Y)-i(U,S)\big].\label{eq:GP_V2}
\end{align}
\end{thm}
\begin{IEEEproof}
See Section \ref{sec:GP_PROOFS}.
\end{IEEEproof}
In the proof of Theorem \ref{thm:GP_MainResult}, we first prove \eqref{eq:GP_MainResult}
with $V$ of the form given in \eqref{eq:GP_V}, and then show that
\eqref{eq:GP_V} and \eqref{eq:GP_V2} coincide under any input distribution
$Q_{U|S}$ which maximizes $I(U;Y)-I(U;S)$ in \eqref{eq:GP_Capacity}.
We will see in Section \ref{sub:GP_EQUIV_VPRIME} that \eqref{eq:GP_V2}
exceeds \eqref{eq:GP_V} more generally; the inequality may be strict
if $Q_{U|S}$ is suboptimal. This is analogous to the single-user
setting, where constant-composition random coding can be used to prove
the achievability of \eqref{eq:GP_SingleUser0} with $V=\mathbb{E}[\mathrm{Var}[i(X,Y)\,|\, X]]$.
This is upper bounded by $\mathrm{Var}[i(X,Y)]$, but equality holds
under any capacity-achieving input distribution \cite{Finite}. 

In Section \ref{sec:GP_CONCLUSION}, we discuss the difficulties in
removing the assumption $\mathbb{E}\big[\mathrm{Var}[i(U,Y)\,|\, S,U]\big]>0$
in Theorem \ref{thm:GP_MainResult}.
\begin{thm}
\label{thm:DPC_MainResult} Consider the dirty paper coding setup
with input power $P>0$. For any sequence of state distributions $P_{\boldsymbol{S}}$
(indexed by $n$) such that
\begin{equation}
\mathbb{P}\big[\|\boldsymbol{S}\|^{2}>n\Pi\big]=O\Big(\frac{\log n}{\sqrt{n}}\Big)\label{eq:DPC_Assumption}
\end{equation}
for some $\Pi<\infty$, we have 
\begin{equation}
\log M^{*}(n,\epsilon)=nC-\sqrt{nV}\mathsf{Q}^{-1}(\epsilon)+O(\log n),\label{eq:DPC_MainResult}
\end{equation}
where
\begin{align}
V & \triangleq\frac{P(2+P)}{2(1+P)^{2}}.\label{eq:DPC_V}
\end{align}
\end{thm}
\begin{IEEEproof}
The converse part follows by revealing the state sequence to the decoder
and using the converse result for the AWGN channel without state \cite{Finite}.
The achievability part is proved in Section \ref{sec:GP_DIRTY_PAPER}.
\end{IEEEproof}
The assumption in \eqref{eq:DPC_Assumption} is mild, allowing for
any state sequence distribution yielding a uniformly bounded (yet
arbitrarily large) power with probability $1-O\big(\frac{\log n}{\sqrt{n}}\big)$.
In particular, the state sequence need not be i.i.d. nor even ergodic,
and may be deterministic. Furthermore, the right-hand of \eqref{eq:DPC_Assumption}
side can be weakened to $o(1)$ at the expense of weakening the $O(\log n)$
term in \eqref{eq:DPC_MainResult} to $o(\sqrt{n})$.

In the case of an i.i.d. state with $S_{i}\sim\pi$, Chebyshev's inequality
reveals that a sufficient condition for \eqref{eq:DPC_Assumption}
to hold is that $\mathbb{E}_{\pi}[S^{4}]<\infty$. In the special
case that $\pi\sim N(0,P_{\pi})$ for some $P_{\pi}>0$, substituting
the capacity-achieving parameters (see \eqref{eq:DPC_OptQ}--\eqref{eq:DPC_OptPhi}
and \eqref{eq:DPC_alpha}) into \eqref{eq:GP_V} yields precisely
the dispersion in \eqref{eq:DPC_V}, thus establishing a connection
with the discrete case. More precisely, the first term in \eqref{eq:GP_V}
equals $\frac{P(2+P)}{2(1+P)^{2}}$, and the second term is zero.

\subsection{Comparisons to Existing Results}

We begin by showing that, for any $\epsilon\in(0,\frac{1}{2})$, Theorem
\ref{thm:GP_MainResult} yields a second-order term which is no worse
than that of \eqref{eq:GP_Existing}, i.e. $\sqrt{V}\mathsf{Q}^{-1}(\epsilon)\le\tilde{R}$.
We claim that
\begin{equation}
\tilde{R}\ge\big(\sqrt{\mathrm{Var}[i(U,S)]}+\sqrt{\mathrm{Var}[i(U,Y)]}\big)\mathsf{Q}^{-1}(\epsilon).
\end{equation}
To see this, we note from \eqref{eq:GP_Qinv2} that any $(\tilde{R}_{1},\tilde{R}_{2})\in\mathsf{Q}_{\mathrm{inv}}(\boldsymbol{V},\epsilon)$
must satisfy $\tilde{R}_{1}\ge\sqrt{V_{1}}\mathsf{Q}^{-1}(\epsilon)$
and $\tilde{R}_{2}\ge\sqrt{V_{2}}\mathsf{Q}^{-1}(\epsilon)$, where
$V_{1}$ and $V_{2}$ are the diagonal entries of $\boldsymbol{V}$.
Furthermore, we can expand \eqref{eq:GP_V2} as 
\begin{align}
V & =\mathrm{Var}[i(U,S)]-2\mathrm{Cov}[i(U,S),i(U,Y)]+\mathrm{Var}[i(U,Y)]\\
 & \le\mathrm{Var}[i(U,S)]+2\sqrt{\mathrm{Var}[i(U,S)]\mathrm{Var}[i(U,Y)]}+\mathrm{Var}[i(U,Y)],\label{eq:GP_Weaken2}
\end{align}
where \eqref{eq:GP_Weaken2} follows from the Cauchy-Schwarz inequality.
The desired result follows from the identity $\sqrt{V_{1}}+\sqrt{V_{2}}=\sqrt{V_{1}+2\sqrt{V_{1}V_{2}}+V_{2}}$,
which is easily verified by squaring both sides.

Next, we present a numerical example showing that the improvement
over \eqref{eq:GP_Existing} can be strict. We revisit the example
of memory with stuck-at faults given in \cite{TanSideInfo}. The alphabets
are given by $\mathcal{S}=\{0,1,2\}$ and $\mathcal{X}=\mathcal{Y}=\{0,1\}$,
and we assume that $\pi=\big(\frac{p}{2},\frac{p}{2},1-p\big)$ for
some constant $p$. The channel is described as follows: $Y=0$ (respectively,
$Y=1$) deterministically whenever $S=0$ (respectively, $S=1$),
and the remaining transition probabilities $W(\cdot|\cdot,2)$ are
those of a binary symmetric channel (BSC) with crossover probability
$\delta$. The capacity is given by
\[
C=(1-p)(1-H_{2}(\delta))\quad\text{bits/use},
\]
 and is achieved by the parameters $\mathcal{U}=\{0,1\}$, $Q_{U|S=0}=(1-\delta,\delta)$,
$Q_{U|S=1}=(\delta,1-\delta)$, $Q_{U|S=2}=\big(\frac{1}{2},\frac{1}{2}\big)$
and $\phi(u,s)=u$. Choosing $\delta=0.11$, $p=0.1$ and $\epsilon=0.001$,
we computed the coefficient to $\sqrt{n}$ in \eqref{eq:GP_Existing}
to be $\tilde{R}=4.16$ $\text{bits}/\sqrt{\text{use}}$, whereas
the coefficient in \eqref{eq:GP_MainResult} is $\sqrt{V}\mathsf{Q}^{-1}(\epsilon)=2.81$
$\text{bits}/\sqrt{\text{use}}$.

The following corollary shows that Theorem \ref{thm:GP_MainResult}
recovers the achievability part of the dispersion for discrete memoryless
channels with state known non-causally at both the encoder and decoder
\cite{TanState}. That is, in this special case there is a matching
converse to Theorem \ref{thm:GP_MainResult}.
\begin{cor}
\label{cor:GP_DecStateInfo}Consider a discrete memoryless state-dependent
channel $W(y|x,s)$ with state distribution $\pi(s)$, where the state
sequence $\boldsymbol{S}$ is known non-causally at both the encoder
and decoder. For each $s\in\mathcal{S}$, let $C_{s}$ and $V_{s}$
denote the capacity and dispersion of the channel $W(\cdot|\cdot,s)$.
If $\mathbb{E}[V_{S}]>0$, then
\begin{equation}
\log M^{*}(n,\epsilon)\ge nC-\sqrt{nV}\mathsf{Q}^{-1}(\epsilon)+O(\log n),
\end{equation}
where
\begin{align}
C & =\mathbb{E}[C_{S}]\\
V & =\mathbb{E}[V_{S}]+\mathrm{Var}[C_{S}].
\end{align}
\end{cor}
\begin{IEEEproof}
See Appendix \ref{sub:GP_DecSIPf}.
\end{IEEEproof}

\section{Proof for the Discrete Memoryless Gel'fand Pinsker Channel \label{sec:GP_PROOFS}}

We present a number of preliminary results in Section \ref{sec:GP_PRELIMINARY},
and we prove Theorem \ref{thm:GP_MainResult} in Section \ref{sub:GP_GENIE_PROOF}.
We make use of the method of types for finite alphabets \cite[Ch. 2]{CsiszarBook}.
The empirical distribution (i.e. type) of a vector $\boldsymbol{x}$
is denoted by $\hat{P}_{\boldsymbol{x}}(x)=\frac{1}{n}\sum_{i=1}^{n}\openone\{x_{i}=x\}$.
The set of all types of length $n$ on an alphabet $\mathcal{X}$
is denoted by $\mathcal{P}_{n}(\mathcal{X})$. The set of all sequences
of length $n$ with a given type $P_{X}$ is denoted by $T^{n}(P_{X})$,
and similarly for joint types.

\subsection{\label{sec:GP_PRELIMINARY}Preliminary Results}

Throughout this subsection, we let $\mathcal{U}$, $Q_{U|S}$ and
$\phi(\cdot,\cdot)$ be arbitrary.

\subsubsection{A Genie-Aided Setting }

We will prove Theorem \ref{thm:GP_MainResult} by first proving the
following result for a genie-aided setting.
\begin{thm}
\label{thm:GP_GenieResult}The statement of Theorem \ref{thm:GP_MainResult}
holds true in the case that the empirical distribution $\hat{P}_{\boldsymbol{S}}$
of $\boldsymbol{S}$ is known at the decoder.
\end{thm}
We proceed by showing that Theorem \ref{thm:GP_GenieResult} implies
Theorem \ref{thm:GP_MainResult}. The idea is to use the first $O(\log n)$
symbols to transmit the type $\hat{P}_{\boldsymbol{S}}$, and then
use a second-order optimal code with state type known at the decoder
for the remaining symbols. This technique was proposed in \cite{GPExponent1}
in the context of random-coding error exponents. 

We fix a sequence $g(n)$, define $\tilde{n}\triangleq n-g(n)$, and
let $P_{S}\in\mathcal{P}_{\tilde{n}}(\mathcal{S})$ be the type of
the last $\tilde{n}$ symbols of $\boldsymbol{S}$. The number of
such types is upper bounded by $\tilde{M}\triangleq(n+1)^{|\mathcal{S}|-1}$.
Using Gallager's random-coding bound \cite[Sec. 5.6]{Gallager}, we
can transmit the type in $g(n)$ symbols with an error probability
$\overline{p}_{e,0}$ satisfying
\begin{equation}
\overline{p}_{e,0}\le e^{-g(n)\tilde{E}_{r}(\tilde{R})},\label{eq:GP_GenieProof1}
\end{equation}
where $\tilde{E}_{r}(\cdot)$ is the random-coding error exponent
of the channel $\tilde{W}(y|x)\triangleq\sum_{s}\pi(s)W(y|x,s)$,
and $\tilde{R}\triangleq\frac{1}{g(n)}\log\tilde{M}=\frac{|\mathcal{S}|-1}{g(n)}\log(n+1)$.%
\footnote{We could instead use the potentially stronger error exponents of \cite{GPExponent1,GPExponent2},
but any exponent which is positive for sufficiently small positive
rates suffices for our purposes.%
} We choose $g(n)=K_{0}\log(n+1)$, where $K_{0}$ is chosen to be
sufficiently large so that $K_{0}\tilde{E}_{r}(\tilde{R})\ge1$ (note
that $\tilde{R}=\frac{|\mathcal{S}|-1}{K_{0}}$, so this is always
possible). It follows from \eqref{eq:GP_GenieProof1} and the choices
of $\tilde{M}$ and $g(n)$ that
\begin{align}
\overline{p}_{e,0} & \le e^{-K_{0}\tilde{E}_{r}(\tilde{R})\log(n+1)}\\
 & \le e^{-\log(n+1)}\\
 & =\frac{1}{n+1}.
\end{align}
Thus, if $\big(n-K_{0}\log(n+1),M,\epsilon-\frac{1}{n+1}\big)$ is
achievable in the genie-aided setting, then $(n,M,\epsilon)$ is achievable
in the absence of the genie. By performing a Taylor expansion of the
square root and $\mathsf{Q}^{-1}(\cdot)$ function in \eqref{eq:GP_MainResult},
we conclude that Theorem \ref{thm:GP_GenieResult} implies Theorem
\ref{thm:GP_MainResult}.

\subsubsection{A Typical Set of State Types}

In Section \ref{sec:GP_PROOFS}, we will study an encoder and decoder
which use different codebooks depending on the type $P_{S}$ of the
state sequence $\boldsymbol{S}$. Here we introduce a typical set
of state types, defined by 
\begin{equation}
\tilde{\mathcal{P}}_{n}\triangleq\bigg\{ P_{S}\in\mathcal{P}_{n}(\mathcal{S})\,:\,\|P_{S}-\pi\|_{\infty}\le\sqrt{\frac{\log n}{n}}\bigg\}.\label{eq:GP_SetPtilde}
\end{equation}
We will see the second-order performance is unaffected by types falling
outside $\tilde{\mathcal{P}}_{n}$, due to the fact that \cite[Lemma 22]{TanState}
\begin{equation}
\mathbb{P}\big[\hat{P}_{\boldsymbol{S}}\notin\tilde{\mathcal{P}}_{n}\big]=O\Big(\frac{1}{n^{2}}\Big).\label{eq:GP_TypicalProb}
\end{equation}
It is clear from \eqref{eq:GP_SetPtilde} that $f(P_{S})\to f(\pi)$
for $P_{S}\in\tilde{\mathcal{P}}_{n}$ and any function $f(\cdot)$
which is continuous at $\pi$, and that $|f(P_{S})-f(\pi)|=O\big(\sqrt{\frac{\log n}{n}}\big)$
for $P_{S}\in\tilde{\mathcal{P}}_{n}$ and any function $f(\cdot)$
which is continuously differentiable at $\pi$.

\subsubsection{Type-Dependent Distributions }

Here we present results regarding the approximation of a distribution
by a type. For each $P_{S}\in\mathcal{P}_{n}(\mathcal{S})$, we define
an approximation $Q_{U|S,n}^{(P_{S})}$ of $Q_{U|S}$ as follows:
For each $s\in\mathcal{S}$ with $P_{S}(s)>0$, let $Q_{U|S,n}^{(P_{S})}(\cdot|s)$
be a type in $\mathcal{P}_{nP_{S}(s)}(\mathcal{U})$ whose probabilities
are $\frac{1}{nP_{S}(s)}$-close to $Q_{U|S}(\cdot|s)$ in terms of
$L_{\infty}$ norm. If $P_{S}(s)=0$ then $Q_{U|S,n}^{(P_{S})}(\cdot|s)$
is arbitrary (e.g. uniform).

Assuming without loss of generality that $\pi(s)>0$ for all $s\in\mathcal{S}$,
we have from \eqref{eq:GP_SetPtilde} that $\min_{s}nP_{S}(s)$ grows
linearly in $n$ for all $P_{S}\in\tilde{\mathcal{P}}_{n}$. It follows
from the above construction of $Q_{U|S,n}^{(P_{S})}$ that 
\begin{equation}
\Big|Q_{U|S}(u|s)-Q_{U|S,n}^{(P_{S})}(u|s)\Big|=O\Big(\frac{1}{n}\Big)\label{eq:GP_Uniform1}
\end{equation}
 uniformly in $P_{S}\in\tilde{\mathcal{P}}_{n}$ and $(s,u)$. 

Throughout Section \ref{sub:GP_GENIE_PROOF}, we will make use of
the following joint distributions:
\begin{align}
P_{SUY}^{(P_{S})}(s,u,y) & \triangleq P_{S}(s)Q_{U|S}(u|s)W(y|\phi(u,s),s)\label{eq:GP_PsDistr}\\
P_{SUY,n}^{(P_{S})}(s,u,y) & \triangleq P_{S}(s)Q_{U|S,n}^{(P_{S})}(u|s)W(y|\phi(u,s),s).\label{eq:GP_PsDistrN}
\end{align}
Using \eqref{eq:GP_Uniform1}, we immediately obtain that 
\begin{equation}
\Big|P_{SUY}^{(P_{S})}(s,u,y)-P_{SUY,n}^{(P_{S})}(s,u,y)\Big|=O\Big(\frac{1}{n}\Big)\label{eq:GP_Uniform2}
\end{equation}
uniformly in $P_{S}\in\tilde{\mathcal{P}}_{n}$ and $(s,u,y)$. We
will use this result to approximate various expectations $\mathbb{E}_{P_{SUY,n}^{(P_{S})}}[\cdot]$
by $\mathbb{E}_{P_{SUY}^{(P_{S})}}[\cdot]$.

\subsubsection{\label{sub:GP_TAYLOR}A Taylor Expansion of the Mutual Information}

Let $I^{(P_{S})}(U;S)$ and $I^{(P_{S})}(U;Y)$ denote mutual informations
with respect to the joint distribution $P_{USY}^{(P_{S})}$ in \eqref{eq:GP_PsDistr},
and define 
\begin{equation}
I(P_{S})\triangleq I^{(P_{S})}(U;Y)-I^{(P_{S})}(U;S).\label{eq:GP_IPS}
\end{equation}
We observe from \eqref{eq:GP_Capacity} that $C=I(\pi)$ whenever
the parameters $\mathcal{U}$, $Q_{U|S}$ and $\phi(\cdot,\cdot)$
achieve capacity. 

In Section \ref{sub:GP_AVG_STATE}, we will make use of a linear approximation
of $I(\cdot)$ given by 
\begin{align}
\tilde{I}(P_{S}) & \triangleq\sum_{s}P_{S}(s)\mathbb{E}[i(U,Y)-i(U,S)\,|\, S=s]\label{eq:GP_Itilde0}\\
 & =\sum_{s}P_{S}(s)\Bigg(\sum_{u,y}Q_{U|S}(u|s)W(y|\phi(u,s),s)\log\frac{P_{Y|U}^{(\pi)}(y|u)}{P_{Y}^{(\pi)}(y)}-\sum_{u}Q_{U|S}(u|s)\log\frac{Q_{U|S}(u|s)}{P_{U}^{(\pi)}(u)}\Bigg),\label{eq:GP_Itilde}
\end{align}
which equals the first-order Taylor approximation of $I(P_{S})$ about
$P_{S}=\pi$. More precisely, we show in Appendix \ref{sub:GP_TAYLOR_EXP}
that
\begin{equation}
I(P_{S})=\tilde{I}(P_{S})+\Delta(P_{S}),\label{eq:GP_Taylor}
\end{equation}
where
\begin{equation}
\max_{P_{S}\in\tilde{\mathcal{P}}_{n}}|\Delta(P_{S})|\le\frac{K_{1}\log n}{n}\label{eq:GP_DeltaProperty}
\end{equation}
for some constant $K_{1}$ and sufficiently large $n$.

\subsection{Proof of Theorem \ref{thm:GP_MainResult} \label{sub:GP_GENIE_PROOF}}

As stated previously, in order to prove Theorem \ref{thm:GP_MainResult},
it suffices to prove Theorem \ref{thm:GP_GenieResult}. Thus, we henceforth
assume that the state type $P_{S}$ is known at the decoder.

\subsubsection{Random-Coding Parameters}

We consider a random-coding ensemble which is similar to that of \cite[Sec. 7.6]{NetworkBook},
the main difference being that we generate a different auxiliary codebook
for each state type. The parameters are the auxiliary alphabet $\mathcal{U}$,
input distribution $Q_{U|S}$, function $\phi\,:\,\mathcal{U}\times\mathcal{S}\to\mathcal{X}$,
and number of auxiliary codewords $L^{(P_{S})}$ per message for each
state type $P_{S}\in\mathcal{P}_{n}(\mathcal{S})$. In accordance
with the statement of Theorem \ref{thm:GP_MainResult}, we assume
that $\mathcal{U}$, $Q_{U|S}$ and $\phi$ are capacity-achieving.

\subsubsection{Codebook Generation}

For each state type $P_{S}\in\mathcal{P}_{n}(\mathcal{S})$ and each
message $m=1,\cdots,M$, we randomly generate an auxiliary codebook
$\mathcal{C}_{U}^{(P_{S})}$ containing $ML^{(P_{S})}$ auxiliary
codewords $\{\boldsymbol{U}^{(P_{S})}(m,l)\}_{l=1}^{L^{(P_{S})}}$,
each of which is independently distributed according to the uniform
distribution on the type class $T^{n}(P_{U,n}^{(P_{S})})$ (see \eqref{eq:GP_PsDistrN}):
\begin{equation}
P_{\boldsymbol{U}}^{(P_{S})}(\boldsymbol{u})=\frac{1}{|T^{n}(P_{U,n}^{(P_{S})})|}\openone\Big\{\boldsymbol{u}\in T^{n}(P_{U,n}^{(P_{S})})\Big\}.\label{eq:GP_PU}
\end{equation}
Each auxiliary codebook is revealed to the encoder and decoder.

\subsubsection{Encoding and Decoding}

Given the state sequence $\boldsymbol{S}\in T^{n}(P_{S})$ and message
$m$, the encoder sends
\begin{equation}
\phi^{n}(\boldsymbol{U},\boldsymbol{S})\triangleq\big(\phi(U_{1},S_{1}),\cdots,\phi(U_{n},S_{n})\big),
\end{equation}
where $\boldsymbol{U}$ is an auxiliary codeword $\boldsymbol{U}^{(P_{S})}(m,l)$
in $\mathcal{C}_{U}^{(P_{S})}$, with $l$ chosen such that $(\boldsymbol{S},\boldsymbol{U})\in T^{n}(P_{SU,n}^{(P_{S})})$.
If multiple such auxiliary codewords exist, one of them is chosen
arbitrarily. An error is declared if no such auxiliary codeword exists.
Given the received vector $\boldsymbol{y}$ and the state type $P_{S}$,
the decoder estimates $m$ according to the pair $(\tilde{m},\tilde{l})$
whose corresponding sequence $\boldsymbol{U}^{(P_{S})}(\tilde{m},\tilde{l})$
maximizes
\begin{equation}
i_{n}^{(P_{S})}(\boldsymbol{u},\boldsymbol{y})\triangleq\sum_{i=1}^{n}i^{(P_{S})}(u_{i},y_{i})\label{eq:GP_iuyMulti}
\end{equation}
among the auxiliary codewords in $\mathcal{C}_{U}^{(P_{S})}$, where
\begin{equation}
i^{(P_{S})}(u,y)\triangleq\log\frac{P_{Y|U}^{(P_{S})}(y|u)}{P_{Y}^{(P_{S})}(y)}\label{eq:GP_iuy_Ps}
\end{equation}
with $P_{SUY}^{(P_{S})}$ defined in \eqref{eq:GP_PsDistr}. Ties
are broken in an arbitrary fashion. Note that $P_{SUY}^{(\pi)}$ coincides
with the joint distribution in \eqref{eq:GP_Distr}, and hence $i^{(\pi)}(u,y)$
coincides with \eqref{eq:GP_iuy}.

We consider the events
\begin{align}
\mathcal{E}_{1} & \triangleq\bigg\{\text{No }l\text{ exists with }(\boldsymbol{S},\boldsymbol{U}^{(P_{S})}(m,l))\in T^{n}(P_{SU,n}^{(P_{S})})\bigg\}\\
\mathcal{E}_{2} & \triangleq\bigg\{\text{Decoder chooses a message }\tilde{m}\ne m\bigg\}.
\end{align}
It follows from these definitions and \eqref{eq:GP_TypicalProb} that
the overall random-coding error probability $\overline{p}_{e}$ satisfies
\begin{equation}
\overline{p}_{e}\le\sum_{P_{S}\in\tilde{\mathcal{P}}_{n}}\mathbb{P}\big[\hat{P}_{\boldsymbol{S}}=P_{S}\big]\Big(\mathbb{P}\big[\mathcal{E}_{1}\,|\, P_{S}\big]+\mathbb{P}\big[\mathcal{E}_{2}\,|\, P_{S},\mathcal{E}_{1}^{c}\big]\Big)+O\Big(\frac{1}{n^{2}}\Big).\label{eq:GP_OverallError}
\end{equation}

\subsubsection{Analysis of $\mathcal{E}_{1}$ \label{sub:GP_ANALYSIS_E1}}

We study the probability of $\mathcal{E}_{1}$ conditioned on $\boldsymbol{S}$
having a given type $P_{S}\in\tilde{\mathcal{P}}_{n}$. Using the
property of types in \cite[Eq. (18)]{GallagerCC}, we have for any
$\boldsymbol{s}\in T^{n}(P_{S})$ and $\boldsymbol{U}$ distributed
according to \eqref{eq:GP_PU} that 
\begin{equation}
\mathbb{P}\big[(\boldsymbol{s},\boldsymbol{U})\in T^{n}(P_{SU,n}^{(P_{S})})\big]\ge\frac{1}{p_{0}(n)}e^{-nI^{(P_{S})}(U;S)},
\end{equation}
where $I^{(P_{S})}(U;S)$ is defined in Section \ref{sub:GP_TAYLOR},
and $p_{0}(n)$ is a polynomial. Since the codewords are generated
independently, it follows that
\begin{align}
\mathbb{P}\big[\mathcal{E}_{1}\,|\, P_{S}\big] & \le\bigg(1-\frac{1}{p_{0}(n)}e^{-nI^{(P_{S})}(U;S)}\bigg)^{L^{(P_{S})}}\\
 & \le\bigg(\exp\bigg(-\frac{1}{p_{0}(n)}e^{-nI^{(P_{S})}(U;S)}\bigg)\bigg)^{L^{(P_{S})}}\label{eq:GP_E1_2}\\
 & =\exp\bigg(-\frac{1}{p_{0}(n)}e^{-n\big(I^{(P_{S})}(U;S)-R_{L}^{(P_{S})}\big)}\bigg),\label{eq:GP_E1_3}
\end{align}
where \eqref{eq:GP_E1_2} follows since $1-\alpha\le e^{-\alpha}$,
and in \eqref{eq:GP_E1_3} we define
\[
R_{L}^{(P_{S})}\triangleq\frac{1}{n}\log L^{(P_{S})}.
\]
Choosing
\begin{equation}
R_{L}^{(P_{S})}=I^{(P_{S})}(U;S)+K_{2}\frac{\log n}{n}\label{eq:GP_ChoiceRL}
\end{equation}
with $K_{2}$ equal to one plus the degree of the polynomial $p_{0}(n)$,
we obtain from \eqref{eq:GP_E1_3} that 
\begin{equation}
\mathbb{P}\big[\mathcal{E}_{1}\,|\, P_{S}\big]\le e^{-\psi n}\label{eq:GP_PE1}
\end{equation}
for some $\psi>0$ and sufficiently large $n$.

\subsubsection{\label{sub:GP_ANALYSIS_E2}Analysis of $\mathcal{E}_{2}$}

We study the probability of $\mathcal{E}_{2}$ conditioned on $\boldsymbol{S}$
having a given type $P_{S}\in\tilde{\mathcal{P}}_{n}$, and also conditioned
on $\mathcal{E}_{1}^{c}$. These conditions imply that $(\boldsymbol{s},\boldsymbol{u})\in T^{n}(P_{SU,n}^{(P_{S})})$
for all $(\boldsymbol{s},\boldsymbol{u})$ occurring with non-zero
probability, and by the symmetry of the state sequence distribution
and the codebook construction, all such $(\boldsymbol{s},\boldsymbol{u})$
are equally likely. It follows that the conditional distribution given
$P_{S}$ and $\mathcal{E}_{1}^{c}$ of the state sequence $\boldsymbol{S}$,
auxiliary codeword $\boldsymbol{U}$, and received sequence $\boldsymbol{Y}$
is given by
\begin{equation}
(\boldsymbol{S},\boldsymbol{U},\boldsymbol{Y})\sim P_{\boldsymbol{S}\boldsymbol{U}}^{(P_{S})}(\boldsymbol{s},\boldsymbol{u})W^{n}(\boldsymbol{y}|\phi^{n}(\boldsymbol{u},\boldsymbol{s}),\boldsymbol{s}),\label{eq:GP_PsVecDistr}
\end{equation}
where
\begin{equation}
P_{\boldsymbol{S}\boldsymbol{U}}^{(P_{S})}(\boldsymbol{s},\boldsymbol{u})\triangleq\frac{1}{\big|T^{n}(P_{SU,n}^{(P_{S})})\big|}\openone\Big\{(\boldsymbol{s},\boldsymbol{u})\in T^{n}(P_{SU,n}^{(P_{S})})\Big\},\label{eq:GP_TypeDistr}
\end{equation}
i.e. the uniform distribution on the type class $T^{n}(P_{SU,n}^{(P_{S})})$.

Recall that the decoder maximizes $i_{n}^{(P_{S})}(\boldsymbol{u},\boldsymbol{y})$
(see \eqref{eq:GP_iuyMulti}) among the $ML^{(P_{S})}$ auxiliary
codewords in $\mathcal{C}_{U}^{(P_{S})}$. Using the threshold bound
for mismatched decoding \cite{SaddlepointThresh}, we have for any
$\gamma^{(P_{S})}$ that
\begin{equation}
\mathbb{P}\big[\mathcal{E}_{2}\,|\, P_{S},\mathcal{E}_{1}^{c}\big]\le\mathbb{P}\Big[i_{n}^{(P_{S})}(\boldsymbol{U},\boldsymbol{Y})\le\gamma^{(P_{S})}\Big]+ML^{(P_{S})}\mathbb{P}\Big[i_{n}^{(P_{S})}(\overline{\boldsymbol{U}},\boldsymbol{Y})>\gamma^{(P_{S})}\Big],\label{eq:GP_ThresholdBound}
\end{equation}
where $\overline{\boldsymbol{U}}\sim P_{\boldsymbol{U}}^{(P_{S})}$
independently of $(\boldsymbol{S},\boldsymbol{U},\boldsymbol{Y})$.
In order to upper bound the second probability, it will prove useful
to upper bound the output distribution $P_{\boldsymbol{Y}}^{(P_{S})}(\boldsymbol{y})\triangleq\sum_{\boldsymbol{s},\boldsymbol{u}}P_{\boldsymbol{S}\boldsymbol{U}}^{(P_{S})}(\boldsymbol{s},\boldsymbol{u})W^{n}(\boldsymbol{y}|\phi^{n}(\boldsymbol{u},\boldsymbol{s}),\boldsymbol{s})$
as follows: 
\begin{align}
P_{\boldsymbol{Y}}^{(P_{S})}(\boldsymbol{y}) & \le p_{1}(n)\sum_{\boldsymbol{s},\boldsymbol{u}}\prod_{i=1}^{n}P_{S}(s_{i})Q_{U|S,n}^{(P_{S})}(u_{i}|s_{i})W(y_{i}|\phi^{n}(u_{i},s_{i}),s_{i})\label{eq:GP_CCtoIID2}\\
 & =p_{1}(n)\prod_{i=1}^{n}\bigg(\sum_{s,u}P_{S}(s)Q_{U|S,n}^{(P_{S})}(u|s)W(y_{i}|\phi(u,s),s)\bigg)\\
 & =p_{1}(n)\prod_{i=1}^{n}\bigg(P_{Y}^{(P_{S})}(y_{i})\Big(1+O\Big(\frac{1}{n}\Big)\Big)\bigg)\label{eq:GP_CCtoIID4}\\
 & \le p_{2}(n)\prod_{i=1}^{n}P_{Y}^{(P_{S})}(y_{i}),\label{eq:GP_CCtoIID5}
\end{align}
where \eqref{eq:GP_CCtoIID2} holds for some polynomial $p_{1}(n)$
by a standard change of measure from uniform on the type class to
i.i.d. \cite[Eq. (2.4)]{Poltyrev}, \eqref{eq:GP_CCtoIID4} follows
from \eqref{eq:GP_Uniform2}, and \eqref{eq:GP_CCtoIID5} follows
for some polynomial $p_{2}(n)$ and sufficiently large $n$ since
$(1+\frac{c}{n})^{n}\to e^{c}$, which is a constant. Using the definition
of $i_{n}^{(P_{S})}$ in \eqref{eq:GP_iuyMulti}, it follows that
\begin{align}
\mathbb{P}\Big[i_{n}^{(P_{S})}(\overline{\boldsymbol{U}},\boldsymbol{Y})>\gamma^{(P_{S})}\Big] & =\sum_{\overline{\boldsymbol{u}},\boldsymbol{y}}P_{\boldsymbol{U}}^{(P_{S})}(\overline{\boldsymbol{u}})P_{\boldsymbol{Y}}^{(P_{S})}(\boldsymbol{y})\openone\bigg\{\prod_{i=1}^{n}\frac{P_{Y|U}^{(P_{S})}(y_{i}|\overline{u}_{i})}{P_{Y}^{(P_{S})}(y_{i})}>e^{\gamma^{(P_{S})}}\bigg\}\label{eq:GP_ThreshAnalysis1a}\\
 & \le\sum_{\overline{\boldsymbol{u}},\boldsymbol{y}}P_{\boldsymbol{U}}^{(P_{S})}(\overline{\boldsymbol{u}})P_{\boldsymbol{Y}}^{(P_{S})}(\boldsymbol{y})\prod_{i=1}^{n}\frac{P_{Y|U}^{(P_{S})}(y_{i}|\overline{u}_{i})}{P_{Y}^{(P_{S})}(y_{i})}e^{-\gamma^{(P_{S})}}\\
 & \le p_{2}(n)\sum_{\overline{\boldsymbol{u}},\boldsymbol{y}}P_{\boldsymbol{U}}^{(P_{S})}(\overline{\boldsymbol{u}})\prod_{i=1}^{n}P_{Y|U}^{(P_{S})}(y_{i}|\overline{u}_{i})e^{-\gamma^{(P_{S})}}\\
 & =p_{2}(n)e^{-\gamma^{(P_{S})}}.\label{eq:GP_ThreshAnalysis1d}
\end{align}
We fix a constant $K_{3}$ and choose
\begin{equation}
\gamma^{(P_{S})}=\log ML^{(P_{S})}+K_{3}\log n\label{eq:GP_ChoiceGamma}
\end{equation}
which, when combined with \eqref{eq:GP_ChoiceRL}, yields 
\begin{equation}
\gamma^{(P_{S})}=\log M+nI^{(P_{S})}(U;S)+K_{4}\log n,\label{eq:GP_Gamma}
\end{equation}
where $K_{4}\triangleq K_{2}+K_{3}$. Setting $K_{3}$ to be one higher
than the degree of $p_{2}(n)$, we obtain from \eqref{eq:GP_ThresholdBound},
\eqref{eq:GP_ThreshAnalysis1d} and \eqref{eq:GP_ChoiceGamma} that
\begin{equation}
ML^{(P_{S})}\mathbb{P}\Big[i_{n}^{(P_{S})}(\overline{\boldsymbol{U}},\boldsymbol{Y})>\gamma^{(P_{S})}\Big]=O\Big(\frac{1}{n}\Big),
\end{equation}
and hence
\begin{equation}
\mathbb{P}\big[\mathcal{E}_{2}\,|\, P_{S},\mathcal{E}_{1}^{c}\big]\le\mathbb{P}\Big[i_{n}^{(P_{S})}(\boldsymbol{U},\boldsymbol{Y})\le\log M+nI^{(P_{S})}(U;S)+K_{4}\log n\Big]+O\Big(\frac{1}{n}\Big),\label{eq:GP_ThreshAnalysis2}
\end{equation}
where the remainder term is uniform in $P_{S}\in\tilde{\mathcal{P}}_{n}$.

\subsubsection{Application of the Berry-Esseen Theorem}

Combining \eqref{eq:GP_PE1} and \eqref{eq:GP_ThreshAnalysis2}, we
have for all $P_{S}\in\tilde{\mathcal{P}}_{n}$ that
\begin{equation}
\mathbb{P}\big[\mathcal{E}_{1}\cup\mathcal{E}_{2}\,|\, P_{S}\big]\le\mathbb{P}\Big[i_{n}^{(P_{S})}(\boldsymbol{U},\boldsymbol{Y})\le\log M+nI^{(P_{S})}(U;S)+K_{4}\log n\Big]+O\Big(\frac{1}{n}\Big).\label{eq:GP_ThreshAnalysis3}
\end{equation}
In order to apply the Berry-Esseen theorem \cite[Sec. XVI.5]{Feller}
to the right-hand side of \eqref{eq:GP_ThreshAnalysis3}, we first
compute the mean and variance of $i_{n}^{(P_{S})}(\boldsymbol{U},\boldsymbol{Y})$,
defined according to \eqref{eq:GP_iuyMulti} and \eqref{eq:GP_PsVecDistr}.
The relevant third moment can easily be uniformly bounded in terms
of the alphabet sizes \cite[Lemma 46]{Finite}, \cite[Appendix D]{MACFinite1}.
We will use the fact that, by the symmetry of the constant-composition
distribution in \eqref{eq:GP_TypeDistr}, the statistics of $i_{n}^{(P_{S})}(\boldsymbol{U},\boldsymbol{Y})$
are unchanged upon conditioning on $(\boldsymbol{S},\boldsymbol{U})=(\boldsymbol{s},\boldsymbol{u})$
for some $(\boldsymbol{s},\boldsymbol{u})\in T^{n}(P_{SU,n}^{(P_{S})})$.
Using the joint distribution $P_{SUY,n}^{(P_{S})}$ defined in \eqref{eq:GP_PsDistrN},
we have
\begin{align}
\mathbb{E}\big[i_{n}^{(P_{S})}(\boldsymbol{u},\boldsymbol{Y})\,|\,\boldsymbol{s},\boldsymbol{u}\big] & =\mathbb{E}\bigg[\sum_{i=1}^{n}i^{(P_{S})}(u_{i},Y_{i})\,\Big|\, s_{i},u_{i}\bigg]\label{eq:GP_Exp1}\\
 & =n\sum_{u,y}P_{UY,n}^{(P_{S})}(u,y)i^{(P_{S})}(u,y)\\
 & =nI^{(P_{S})}(U;Y)+O(1),\label{eq:GP_Exp2}
\end{align}
where \eqref{eq:GP_Exp2} follows from \eqref{eq:GP_Uniform2} and
the definitions of $i^{(P_{S})}(u,y)$ and $I^{(P_{S})}(U;Y)$ (see
\eqref{eq:GP_iuy_Ps} and Section \ref{sub:GP_TAYLOR}). Similarly,
we have
\begin{align}
\mathrm{Var}\big[i_{n}^{(P_{S})}(\boldsymbol{u},\boldsymbol{Y})\,|\,\boldsymbol{s},\boldsymbol{u}\big] & =\mathrm{Var}\bigg[\sum_{i=1}^{n}i^{(P_{S})}(u_{i},Y_{i})\,\Big|\, s_{i},u_{i}\bigg]\\
 & =n\sum_{s,u}P_{SU,n}^{(P_{S})}(s,u)\mathrm{Var}\big[i^{(P_{S})}(u,Y)\,|\, s,u\big]\label{eq:GP_VarPs}\\
 & =n\mathbb{E}\Big[\mathrm{Var}\big[i^{(P_{S})}(U,Y)\,|\, S,U\big]\Big]+O(1)\\
 & \triangleq nV(P_{S})+O(1).\label{eq:GP_Var2}
\end{align}
It should be noted that $V(P_{S})$ is bounded away for zero for $P_{S}\in\tilde{\mathcal{P}}_{n}$
and sufficiently large $n$, since $V(\pi)>0$ by assumption in Theorem
\ref{thm:GP_MainResult}. Furthermore, the $O(1)$ terms in \eqref{eq:GP_Exp2}
and \eqref{eq:GP_Var2} are uniform in $P_{S}\in\tilde{\mathcal{P}}_{n}$,
due to the uniformity of \eqref{eq:GP_Uniform2}.

Using the definition of $I(P_{S})$ in \eqref{eq:GP_IPS}, we choose
\begin{equation}
\log M=nI(\pi)-K_{4}\log n-\beta_{n},\label{eq:GP_ChoiceM}
\end{equation}
where $\beta_{n}$ will be specified later, and will behave as $O(\sqrt{n})$.
Combining \eqref{eq:GP_ThreshAnalysis3}, \eqref{eq:GP_Exp2}, \eqref{eq:GP_Var2}
and \eqref{eq:GP_ChoiceM}, we have 
\begin{align}
\mathbb{P}\big[\mathcal{E}_{1}\cup\mathcal{E}_{2}\,|\, P_{S}\big] & \le\mathbb{P}\Big[i_{n}^{(P_{S})}(\boldsymbol{U},\boldsymbol{Y})\le nI(\pi)+nI^{(P_{S})}(U;S)-\beta_{n}\Big]+O\Big(\frac{1}{n}\Big).\\
 & =\mathbb{P}\Big[i_{n}^{(P_{S})}(\boldsymbol{u},\boldsymbol{Y})\le nI(\pi)+nI^{(P_{S})}(U;S)-\beta_{n}\,\big|\,\boldsymbol{s},\boldsymbol{u}\Big]+O\Big(\frac{1}{n}\Big)\label{eq:GP_ThreshAnalysis4}\\
 & \le\mathsf{Q}\bigg(\frac{\beta_{n}+nI(P_{S})-nI(\pi)+K_{5}}{\sqrt{nV(P_{S})+K_{6}}}\bigg)+O\bigg(\frac{1}{\sqrt{n}}\bigg)\label{eq:GP_ThreshAnalysis5}
\end{align}
where \eqref{eq:GP_ThreshAnalysis4} holds for any $(\boldsymbol{s},\boldsymbol{u})\in T^{n}(P_{SU,n}^{(P_{S})})$
by symmetry, and \eqref{eq:GP_ThreshAnalysis5} follows from the Berry-Esseen
theorem for independent and non-identically distributed variables
\cite[Sec. XVI.5]{Feller}, and by introducing the constants $K_{5}$
and $K_{6}$ to represent the uniform $O(1)$ terms in \eqref{eq:GP_Exp2}
and \eqref{eq:GP_Var2}.

\subsubsection{\label{sub:GP_AVG_STATE}Averaging Over the State Type}

Substituting \eqref{eq:GP_ThreshAnalysis5} into \eqref{eq:GP_OverallError},
we have
\begin{align}
\overline{p}_{e} & \le\sum_{P_{S}\in\tilde{\mathcal{P}}_{n}}\mathbb{P}\big[\hat{P}_{\boldsymbol{S}}=P_{S}\big]\mathsf{Q}\bigg(\frac{\beta_{n}+nI(P_{S})-nI(\pi)+K_{5}}{\sqrt{nV(P_{S})+K_{6}}}\bigg)+O\Big(\frac{1}{\sqrt{n}}\Big)\label{eq:GP_AvgOverS}\\
 & \le\sum_{P_{S}\in\tilde{\mathcal{P}}_{n}}\mathbb{P}\big[\hat{P}_{\boldsymbol{S}}=P_{S}\big]\mathsf{Q}\bigg(\frac{\beta_{n}+nI(P_{S})-nI(\pi)}{\sqrt{nV(P_{S})}}\bigg)+O\Big(\frac{1}{\sqrt{n}}\Big),\label{eq:GP_AvgOverS2}
\end{align}
where \eqref{eq:GP_AvgOverS2} holds for any $\beta_{n}=O(\sqrt{n})$
using standard inequalities based on Taylor expansions; see Appendix
\ref{sub:GP_PROOFS_STEPS} for details. Analogously to \cite[Lemmas 17-18]{TanState},
we simplify \eqref{eq:GP_AvgOverS2} using two lemmas.
\begin{lem}
\label{lem:GP_Lem1} For any $\beta_{n}=O(\sqrt{n})$, we have
\begin{equation}
\sum_{P_{S}\in\tilde{\mathcal{P}}_{n}}\mathbb{P}\big[\hat{P}_{\boldsymbol{S}}=P_{S}\big]\mathsf{Q}\Bigg(\frac{\beta_{n}+nI(P_{S})-nI(\pi)}{\sqrt{nV(P_{S})}}\Bigg)\le\sum_{P_{S}\in\tilde{\mathcal{P}}_{n}}\mathbb{P}\big[\hat{P}_{\boldsymbol{S}}=P_{S}\big]\mathsf{Q}\Bigg(\frac{\beta_{n}+nI(P_{S})-nI(\pi)}{\sqrt{nV(\pi)}}\Bigg)+O\Big(\frac{\log n}{\sqrt{n}}\Big)\label{eq:GP_Lem1}
\end{equation}
\end{lem}
\begin{IEEEproof}
Since $V(P_{S})$ is continuously differentiable at $P_{S}=\pi$ (see
Appendix \ref{sub:GP_TAYLOR_EXP}), a Taylor expansion and the definition
of $\tilde{\mathcal{P}}_{n}$ in \eqref{eq:GP_SetPtilde} yields that
the left-hand side of \eqref{eq:GP_Lem1} is upper bounded by
\begin{equation}
\sum_{P_{S}\in\tilde{\mathcal{P}}_{n}}\mathbb{P}\big[\hat{P}_{\boldsymbol{S}}=P_{S}\big]\mathsf{Q}\Bigg(\frac{\beta_{n}+nI(P_{S})-nI(\pi)}{\sqrt{n\big(V(\pi)+K_{7}\sqrt{\frac{\log n}{n}}\big)}}\Bigg)\label{eq:GP_Lem1_2}
\end{equation}
for some constant $K_{7}$. In Appendix \ref{sub:GP_PROOFS_STEPS},
we show that \eqref{eq:GP_Lem1_2} is upper bounded by the right-hand
side of \eqref{eq:GP_Lem1} using the assumption $\beta_{n}=O(\sqrt{n})$
along with standard inequalities based on Taylor expansions.
\end{IEEEproof}
Lemma \ref{lem:GP_Lem1} is analogous to \cite[Lemma 17]{TanState},
which is proved in a different manner using Hermite polynomials. The
proof in \cite{TanState} is somewhat more involved than that of Lemma
\ref{lem:GP_Lem1}, but it does not make the assumption that $\beta_{n}=O(\sqrt{n})$,
and it yields a tighter $O\big(\frac{\log n}{n}\big)$ remainder term.
\begin{lem}
\label{lem:GP_Lem2} For any $\beta_{n}$, we have
\begin{equation}
\sum_{P_{S}\in\tilde{\mathcal{P}}_{n}}\mathbb{P}\big[\hat{P}_{\boldsymbol{S}}=P_{S}\big]\mathsf{Q}\bigg(\frac{\beta_{n}+nI(P_{S})-nI(\pi)}{\sqrt{nV(\pi)}}\bigg)\le\mathsf{Q}\bigg(\frac{\beta_{n}}{\sqrt{nV}}\bigg)+O\bigg(\frac{\log n}{\sqrt{n}}\bigg),\label{eq:GP_Lem2}
\end{equation}
where $V$ is defined in \eqref{eq:GP_V}.\end{lem}
\begin{IEEEproof}
Using the expansion of $I(P_{S})$ in terms of $\tilde{I}(P_{S})$
and $\Delta(P_{S})$ given in \eqref{eq:GP_Taylor}, we have
\begin{align}
 & \sum_{P_{S}\in\tilde{\mathcal{P}}_{n}}\mathbb{P}\big[\hat{P}_{\boldsymbol{S}}=P_{S}\big]\mathsf{Q}\bigg(\frac{\beta_{n}+nI(P_{S})-nI(\pi)}{\sqrt{nV(\pi)}}\bigg)\\
 & \qquad=\sum_{P_{S}\in\tilde{\mathcal{P}}_{n}}\mathbb{P}\big[\hat{P}_{\boldsymbol{S}}=P_{S}\big]\mathsf{Q}\Bigg(\frac{\beta_{n}-nI(\pi)+n\tilde{I}(P_{S})+n\Delta(P_{S})}{\sqrt{nV(\pi)}}\Bigg)\\
 & \qquad\le\sum_{P_{S}\in\tilde{\mathcal{P}}_{n}}\mathbb{P}\big[\hat{P}_{\boldsymbol{S}}=P_{S}\big]\mathsf{Q}\Bigg(\frac{\beta_{n}-nI(\pi)+n\tilde{I}(P_{S})-2K_{1}\log n}{\sqrt{nV(\pi)}}\Bigg)\label{eq:GP_Lem2_5}\\
 & \qquad=\sum_{P_{S}\in\tilde{\mathcal{P}}_{n}}\mathbb{P}\big[\hat{P}_{\boldsymbol{S}}=P_{S}\big]\mathsf{Q}\Bigg(\frac{\beta_{n}-nI(\pi)+n\tilde{I}(P_{S})}{\sqrt{nV(\pi)}}\Bigg)+O\Big(\frac{\log n}{\sqrt{n}}\Big)\label{eq:GP_Lem2_6}\\
 & \qquad\le\sum_{P_{S}}\mathbb{P}\big[\hat{P}_{\boldsymbol{S}}=P_{S}\big]\mathsf{Q}\Bigg(\frac{\beta_{n}-nI(\pi)+n\tilde{I}(P_{S})}{\sqrt{nV(\pi)}}\Bigg)+O\Big(\frac{\log n}{\sqrt{n}}\Big),\label{eq:GP_Lem2_7}
\end{align}
where \eqref{eq:GP_Lem2_5} follows from \eqref{eq:GP_DeltaProperty}
and since $\mathsf{Q}(\cdot)$ is decreasing, and \eqref{eq:GP_Lem2_6}
follows from the identity $|\mathsf{Q}(z)-\mathsf{Q}(z+a)|\le\frac{|a|}{\sqrt{2\pi}}$.
Since $\tilde{I}(P_{S})$ is written in the form $\sum_{s}P_{S}(s)\psi(s)$,
a trivial generalization of \cite[Lemma 18]{TanState} gives 
\begin{equation}
\sum_{P_{S}}\mathbb{P}\big[\hat{P}_{\boldsymbol{S}}=P_{S}\big]\mathsf{Q}\Bigg(\frac{\beta_{n}+n\tilde{I}(P_{S})-n\tilde{I}(\pi)}{\sqrt{nV(\pi)}}\Bigg)-\mathsf{Q}\Bigg(\frac{\beta_{n}}{\sqrt{n\big(V(\pi)+V^{*}(\pi)\big)}}\Bigg)=O\Big(\frac{1}{\sqrt{n}}\Big),\label{eq:GP_Lem2_8}
\end{equation}
where $V^{*}(\pi)\triangleq\mathrm{Var}_{\pi}[\psi(S)]$. Using the
definition of $V(\cdot)$ in \eqref{eq:GP_Var2} and the fact that
$\psi(S)=\mathbb{E}[i(U,Y)-i(U,S)\,|\, S]$ (see \eqref{eq:GP_Itilde0}),
it follows that $V(\pi)+V^{*}(\pi)$ is equal to $V$, defined in
\eqref{eq:GP_V}.
\end{IEEEproof}
Using \eqref{eq:GP_AvgOverS2} along with Lemmas \ref{lem:GP_Lem1}
and \ref{lem:GP_Lem2}, we have
\begin{equation}
\overline{p}_{e}\le\mathsf{Q}\bigg(\frac{\beta_{n}}{\sqrt{nV}}\bigg)+O\Big(\frac{\log n}{\sqrt{n}}\Big).
\end{equation}
Setting $\overline{p}_{e}=\epsilon$ and solving for $\beta_{n}$,
we obtain
\begin{equation}
\beta_{n}=\sqrt{nV}\mathsf{Q}^{-1}(\epsilon)+O(\log n).\label{eq:GP_Betan}
\end{equation}
Consistent with the step \eqref{eq:GP_AvgOverS2} and the statement
of Lemma \ref{lem:GP_Lem1}, we have $\beta_{n}=O(\sqrt{n})$. Finally,
substituting \eqref{eq:GP_Betan} into \eqref{eq:GP_ChoiceM}, we
obtain \eqref{eq:GP_MainResult}.

\subsubsection{\label{sub:GP_EQUIV_VPRIME}Equivalent form of $V$}

It remains to show that \eqref{eq:GP_V2} holds. We have
\begin{align}
\mathrm{Var}[i(U,Y)-i(U,S)] & =\mathbb{E}\big[\mathrm{Var}[i(U,Y)-i(U,S)\,|\, S,U]\big]+\mathrm{Var}\big[\mathbb{E}[i(U,Y)-i(U,S)\,|\, S,U]\big]\label{eq:GP_Weaken1}\\
 & =\mathbb{E}\big[\mathrm{Var}[i(U,Y)\,|\, S,U]\big]+\mathrm{Var}\big[\mathbb{E}[i(U,Y)-i(U,S)\,|\, S,U]\big]\\
 & \ge\mathbb{E}\big[\mathrm{Var}[i(U,Y)\,|\, S,U]\big]+\mathrm{Var}\big[\mathbb{E}[i(U,Y)-i(U,S)\,|\, S]\big],\label{eq:GP_Weaken3}
\end{align}
where \eqref{eq:GP_Weaken1} follows from the law of total variance,
and \eqref{eq:GP_Weaken3} follows by again using the law of total
variance to write
\begin{equation}
\mathbb{E}\big[\mathrm{Var}[\cdot|S,U]\big]+\mathrm{Var}\big[\mathbb{E}[\cdot|S,U]\big]=\mathbb{E}\big[\mathrm{Var}[\cdot|S]\big]+\mathrm{Var}\big[\mathbb{E}[\cdot|S]\big],
\end{equation}
and since $\mathbb{E}\big[\mathrm{Var}[\cdot|\, S,U]\big]\le\mathbb{E}\big[\mathrm{Var}[\cdot|\, S]\big]$.
We show in Appendix \ref{sub:GP_NECC_CONDS} that whenever $Q_{U|S}$
maximizes the objective in \eqref{eq:GP_Capacity}, we have for any
$s\in\mathcal{S}$ that the quantity $\xi^{\prime}(s,u)\triangleq\mathbb{E}[i(u,Y)-i(u,s)\,|\, s,u]$
takes a fixed value $\xi(s)$ for all $u$ such that $Q_{U|S}(u|s)>0$.
It follows that $\mathrm{Var}[\xi^{\prime}(S,U)]=\mathrm{Var}[\xi(S)]$,
and hence \eqref{eq:GP_Weaken3} holds with equality. Since we are
considering capacity-achieving parameters, we obtain \eqref{eq:GP_V2},
thus completing the proof of Theorem \ref{thm:GP_MainResult}.

\section{Proof for Dirty Paper Coding \label{sec:GP_DIRTY_PAPER}}

In this section, we prove Theorem \ref{thm:DPC_MainResult} by adapting
the analysis of Section \ref{sec:GP_PROOFS} to the Gaussian setting.
To highlight the similarities in the proofs, we use similar or identical
notation for analogous quantities.

\subsection{Preliminary Results}

\subsubsection{Power Types}

In place of types based on empirical distributions, we make use of
\emph{power types }(e.g. see \cite{UniversalGaussian}). We fix $\delta_{s}>0$,
and for each $P_{S}=\frac{k\delta_{s}}{n}$ ($k=0,1,2,\cdots$), we
define the type class
\begin{align}
T^{n}(P_{S}) & \triangleq\Big\{\boldsymbol{s}\,:\, nP_{S}\le\|\boldsymbol{s}\|^{2}<nP_{S}+\delta_{s}\Big\}.\label{eq:DPC_SetTn}
\end{align}
For each $\boldsymbol{s}\in T^{n}(P_{S})$, we say that $P_{S}$ is
the type of $\boldsymbol{s}$, and we write $\hat{P}_{\boldsymbol{s}}=P_{S}$.
That is, the type of a sequence is its power rounded down to the nearest
multiple of $\frac{\delta_{s}}{n}$. The set of all types is given
by $\mathcal{P}_{n}\triangleq\big\{\frac{k\delta_{s}}{n}\,:\, k\in\mathbb{Z}\big\}$.

\subsubsection{A Typical Set of State Types}

In general, the type $P_{S}$ of $\boldsymbol{S}$ can be arbitrarily
large with non-zero probability. However, analogously to \eqref{eq:GP_SetPtilde},
we can define a typical set of state types as follows: 
\begin{equation}
\tilde{\mathcal{P}}_{n}\triangleq\big\{ P_{S}\in\mathcal{P}_{n}\,:\, P_{S}\le\Pi\big\}.\label{eq:DPC_TypicalSet}
\end{equation}
where $\Pi$ appears in \eqref{eq:DPC_Assumption}. We immediately
obtain from \eqref{eq:DPC_Assumption} that 
\begin{equation}
\mathbb{P}\big[\hat{P}_{\boldsymbol{S}}\notin\tilde{\mathcal{P}}_{n}\big]\le O\Big(\frac{\log n}{\sqrt{n}}\Big).\label{eq:DPC_TypicalProb}
\end{equation}
Furthermore, the number of state types falling into $\tilde{\mathcal{P}}_{n}$
grows as $\Theta(n)$.

\subsubsection{A Genie-Aided Setting}

Analogously to the discrete case, we will prove Theorem \ref{thm:DPC_GenieResult}
via the following result for a genie-aided setting.
\begin{thm}
\label{thm:DPC_GenieResult}The statement of Theorem \ref{thm:DPC_MainResult}
holds true in the case that the type $P_{S}$ of $\boldsymbol{S}$
is known at the decoder.
\end{thm}
We proceed by showing that Theorem \ref{thm:DPC_GenieResult} implies
Theorem \ref{thm:DPC_MainResult}. The arguments are similar to those
following Theorem \ref{thm:GP_GenieResult}, so we only state the
differences. We treat the event $\hat{P}_{\boldsymbol{S}}\notin\tilde{\mathcal{P}}_{n}$
as an error, thus leaving one of $\Theta(n)$ types to be transmitted
to the receiver in $O(\log n)$ channel uses. This can be done provided
that we can find a random-coding error exponent which is positive
for sufficiently small rates. That is, we wish to show that there
exists $\delta>0$ and $\psi>0$ such that for sufficiently large
$n$ the error probability does does not exceed $e^{-n\psi}$ for
$R\le\delta$.

From \cite[Prop. 1]{ExpAVC}, a positive exponent can be achieved
for rates below $\frac{1}{2}\log\big(1+\frac{P}{1+P_{\mathrm{max}}}\big)$
even when the state sequence $\boldsymbol{S}$ is unknown at the encoder
and arbitrarily varying subject to $\|\boldsymbol{S}\|^{2}\le nP_{\mathrm{max}}$.
Since we have treated the event $\hat{P}_{\boldsymbol{S}}\notin\tilde{\mathcal{P}}_{n}$
as an error, it follows from \eqref{eq:DPC_TypicalSet} that the desired
exponential decay is achieved for rates below $\frac{1}{2}\log\big(1+\frac{P}{1+\Pi}\big)$.

\subsubsection{Type-Dependent Distributions}

We will consider a decoder which makes use of an information density
defined with respect to the joint distribution
\begin{equation}
f_{SUY}^{(P_{S})}(s,u,y)=f_{S}^{(P_{S})}(s)Q_{U|S}(u|s)f_{Y|SU}(y|s,u),\label{eq:DPC_fSUY}
\end{equation}
where in accordance with \eqref{eq:DPC_Channel} and \eqref{eq:DPC_OptQ}--\eqref{eq:DPC_OptPhi},
we have 
\begin{align}
f_{S}^{(P_{S})} & \sim N(0,P_{S})\label{eq:DPC_fS}\\
Q_{U|S} & \sim N(-\alpha S,P)\label{eq:DPC_fU|S}\\
f_{Y|SU} & \sim N\big(U+(1-\alpha)S,1\big).\label{eq:DPC_fY|SU}
\end{align}
The parameter $\alpha>0$ is assumed to be arbitrary for now. We can
think of $f_{SUY}^{(P_{S})}$ as being the joint density of $(S,U,Y)$
induced by a Gaussian state $S\sim N(0,P_{S})$, the channel $W$,
and the choices of $Q_{U|S}$ and $\phi$ in \eqref{eq:DPC_OptQ}--\eqref{eq:DPC_OptPhi}.
This joint density will play a major role in the analysis even though
we are considering a possibly non-Gaussian state sequence. The induced
output distribution is given by
\begin{equation}
f_{Y}^{(P_{S})}\sim N(0,P+P_{S}+1),\label{eq:DPC_fY}
\end{equation}
and similarly to \eqref{eq:DPC_IUY}--\eqref{eq:DPC_IUS}, the corresponding
mutual informations are given by
\begin{align}
I^{(P_{S})}(U;Y) & =\frac{1}{2}\log\Bigg(\frac{(P+P_{S}+1)(P+\alpha^{2}P_{S})}{PP_{S}(1-\alpha)^{2}+(P+\alpha^{2}P_{S})}\Bigg)\label{eq:DPC_IUY_Ps}\\
I^{(P_{S})}(U;S) & =\frac{1}{2}\log\Bigg(\frac{P+\alpha^{2}P_{S}}{P}\Bigg).\label{eq:DPC_IUS_Ps}
\end{align}

\subsection{Proof of Theorem \ref{thm:DPC_MainResult}}

As stated previously, in order to prove Theorem \ref{thm:DPC_MainResult},
it suffices to prove Theorem \ref{thm:DPC_GenieResult}. Thus, we
henceforth assume that the state type $P_{S}$ is known at the decoder.

\subsubsection{Random-Coding Parameters}

The random coding parameters are the constant $\alpha>0$ and the
number of auxiliary codewords for each state type $P_{S}\in\mathcal{P}_{n}$,
denoted by $L^{(P_{S})}$. We will perform the analysis for an arbitrary
choice of $\alpha>0$, and then substitute $\alpha=\frac{P}{1+P}$
in accordance with \eqref{eq:DPC_alpha}.

\subsubsection{Codebook Generation}

For each state type $P_{S}\in\mathcal{P}_{n}$ and each message $m$,
we randomly generate an auxiliary codebook $\mathcal{C}_{U}^{(P_{S})}$
containing $ML^{(P_{S})}$ auxiliary codewords $\{\boldsymbol{U}^{(P_{S})}(m,l)\}_{l=1}^{L^{(P_{S})}}$,
where each codeword is independently distributed according to the
uniform distribution on the sphere of power $n(P+\alpha^{2}P_{S})$,
namely
\begin{equation}
f_{\boldsymbol{U}}^{(P_{S})}(\boldsymbol{u})=\frac{\delta\big(\|\boldsymbol{u}\|^{2}-n(P+\alpha^{2}P_{S})\big)}{S_{n}\big(\sqrt{n(P+\alpha^{2}P_{S})}\big)},\label{eq:DPC_AuxDistr}
\end{equation}
where $\delta(\cdot)$ is the Dirac delta function, and
\begin{equation}
S_{n}(r)=\frac{2\pi^{\frac{n}{2}}}{\Gamma\big(\frac{n}{2}\big)}r^{n-1}\label{eq:DPC_Sn}
\end{equation}
 is the surface area of a sphere of radius $r$ in $n$-dimensional
space. Each auxiliary codebook is revealed to the encoder and decoder.

\subsubsection{Encoding and Decoding}

Given the state sequence $\boldsymbol{S}\in T^{n}(P_{S})$ and message
$m$, the encoder sends
\begin{equation}
\boldsymbol{X}=\boldsymbol{U}-\alpha\boldsymbol{S},
\end{equation}
where $\boldsymbol{U}$ is an auxiliary codeword $\boldsymbol{U}^{(P_{S})}(m,l)$
in $\mathcal{C}_{U}^{(P_{S})}$, with $l$ chosen such that $\boldsymbol{X}\in\mathcal{D}_{n}$,
where 
\begin{equation}
\mathcal{D}_{n}\triangleq\Big\{\boldsymbol{x}\,:\, nP-\delta_{x}\le\|\boldsymbol{x}\|^{2}\le nP\Big\}\label{eq:DPC_Xcond}
\end{equation}
for some $\delta_{x}>0$. If multiple such auxiliary codewords exist,
one of them is chosen arbitrarily. An error is declared if no such
auxiliary codeword exists. By construction, the power constraint in
\eqref{eq:DPC_InputConstr} is satisfied with probability one.

Given the received vector $\boldsymbol{y}$ and the state type $P_{S}$,
the decoder estimates $m$ according to the pair $(\tilde{m},\tilde{l})$
whose corresponding sequence $\boldsymbol{U}^{(P_{S})}(\tilde{m},\tilde{l})$
maximizes
\begin{equation}
i_{n}^{(P_{S})}(\boldsymbol{u},\boldsymbol{y})\triangleq\sum_{i=1}^{n}i^{(P_{S})}(u_{i},y_{i})\label{eq:GP_iuyMulti-1}
\end{equation}
among the auxiliary codewords in $\mathcal{C}_{U}^{(P_{S})},$ where
\begin{equation}
i^{(P_{S})}(u,y)\triangleq\log\frac{f_{Y|U}^{(P_{S})}(y|u)}{f_{Y}^{(P_{S})}(y)}\label{eq:DPC_iuy}
\end{equation}
with $f_{SUY}^{(P_{S})}$ defined in \eqref{eq:DPC_fSUY}.

We consider the events
\begin{align}
\mathcal{E}_{1} & \triangleq\bigg\{\text{No }l\text{ exists such that }\boldsymbol{U}^{(P_{S})}(m,l)-\alpha\boldsymbol{S}\in\mathcal{D}_{n}\bigg\}\\
\mathcal{E}_{2} & \triangleq\bigg\{\text{Decoder chooses a message }\tilde{m}\ne m\bigg\}.
\end{align}
It follows from these definitions and \eqref{eq:DPC_TypicalProb}
that the overall random-coding error probability $\overline{p}_{e}$
satisfies
\begin{equation}
\overline{p}_{e}\le\sum_{P_{S}\in\tilde{\mathcal{P}}_{n}}\mathbb{P}\big[\hat{P}_{\boldsymbol{S}}=P_{S}\big]\Big(\mathbb{P}\big[\mathcal{E}_{1}\,|\, P_{S}\big]+\mathbb{P}\big[\mathcal{E}_{2}\,|\, P_{S},\mathcal{E}_{1}^{c}\big]\Big)+O\Big(\frac{\log n}{\sqrt{n}}\Big).\label{eq:DPC_OverallError}
\end{equation}

\subsubsection{Analysis of $\mathcal{E}_{1}$}

We study the probability of $\mathcal{E}_{1}$ conditioned on $\boldsymbol{S}$
having a given type $P_{S}\in\tilde{\mathcal{P}}_{n}$. Recall the
definition of $I^{(P_{S})}(U;S)$ in \eqref{eq:DPC_IUS_Ps}. We claim
that there exists a constant $K_{1}$ such that the rate $R_{L}^{(P_{S})}\triangleq\frac{1}{n}\log L^{(P_{S})}$
can be set to 
\begin{equation}
R_{L}^{(P_{S})}=I^{(P_{S})}(U;S)+K_{1}\frac{\log n}{n}\label{eq:DPC_RL}
\end{equation}
while achieving
\begin{equation}
\mathbb{P}\big[\mathcal{E}_{1}\,|\, P_{S}\big]\le e^{-\psi n}\label{eq:DPC_E1decay}
\end{equation}
for some $\psi>0$ and sufficiently large $n$. The key result in
proving this claim is the following.
\begin{lem}
\label{lem:DPC_Geometric}Fix $P_{S}\in\tilde{\mathcal{P}}_{n}$,
and let $\boldsymbol{U}$ have density $f_{\boldsymbol{U}}^{(P_{S})}$
(see \eqref{eq:DPC_AuxDistr}). For all $\boldsymbol{s}\in T^{n}(P_{S})$
and sufficiently large $n$, we have
\begin{equation}
\mathbb{P}\big[\boldsymbol{U}-\alpha\boldsymbol{s}\in\mathcal{D}_{n}\big]\ge\frac{1}{p_{0}(n)}e^{-I^{(P_{S})}(U;Y)},\label{eq:DPC_Geometric}
\end{equation}
for some polynomial $p_{0}(n)$ not depending on $P_{S}$.\end{lem}
\begin{IEEEproof}
See Appendix \ref{sub:DPC_PF_GEOMETRIC}.
\end{IEEEproof}
We obtain \eqref{eq:DPC_E1decay} using Lemma \ref{lem:DPC_Geometric}
and following identical steps to those given in Section \ref{sub:GP_ANALYSIS_E1};
the remaining details are omitted to avoid repetition.

\subsubsection{Analysis of $\mathcal{E}_{2}$}

We study the probability of $\mathcal{E}_{2}$ conditioned on $\boldsymbol{S}$
having a given type $P_{S}\in\tilde{\mathcal{P}}_{n}$, and also conditioned
on $\mathcal{E}_{1}^{c}$. Let $f_{\boldsymbol{S}\boldsymbol{U}}^{(P_{S})}(\boldsymbol{s},\boldsymbol{u})$
denote the joint density of $(\boldsymbol{S},\boldsymbol{U})$ conditioned
on these events, and let $\boldsymbol{Y}$ by the resulting output
random variable, i.e.
\begin{equation}
(\boldsymbol{S},\boldsymbol{U},\boldsymbol{Y})\sim f_{\boldsymbol{S}\boldsymbol{U}}^{(P_{S})}(\boldsymbol{s},\boldsymbol{u})W^{n}(\boldsymbol{y}|\boldsymbol{u}-\alpha\boldsymbol{s},\boldsymbol{s}).\label{eq:DPC_CondDistr}
\end{equation}
We do not attempt to give an explicit characterization of $f_{\boldsymbol{S}\boldsymbol{U}}^{(P_{S})}$.
Instead, we will derive properties of the distribution which will
be sufficient for performing the analysis; see Lemmas \ref{lem:DPC_Property1}
and \ref{lem:DPC_Property2} below. 

We again use the threshold-based bound given in \eqref{eq:GP_ThresholdBound},
which states that
\begin{equation}
\mathbb{P}\big[\mathcal{E}_{2}\,|\, P_{S},\mathcal{E}_{1}^{c}\big]\le\mathbb{P}\Big[i_{n}^{(P_{S})}(\boldsymbol{U},\boldsymbol{Y})\le\gamma^{(P_{S})}\Big]+ML^{(P_{S})}\mathbb{P}\Big[i_{n}^{(P_{S})}(\overline{\boldsymbol{U}},\boldsymbol{Y})>\gamma^{(P_{S})}\Big]\label{eq:DPC_ThresholdBound}
\end{equation}
for any $\gamma^{(P_{S})}$, where $\overline{\boldsymbol{U}}\sim f_{\boldsymbol{U}}^{(P_{S})}$
is independent of $(\boldsymbol{S},\boldsymbol{U},\boldsymbol{Y})$.
We further upper bound \eqref{eq:DPC_ThresholdBound} by maximizing
over $(\boldsymbol{s},\boldsymbol{u})$:
\begin{equation}
\mathbb{P}\big[\mathcal{E}_{2}\,|\, P_{S},\mathcal{E}_{1}^{c}\big]\le\max_{(\boldsymbol{s},\boldsymbol{u})\,:\, f_{\boldsymbol{S}\boldsymbol{U}}^{(P_{S})}(\boldsymbol{s},\boldsymbol{u})>0}\mathbb{P}\Big[i_{n}^{(P_{S})}(\boldsymbol{u},\boldsymbol{Y})\le\gamma^{(P_{S})}\,\Big|\,\boldsymbol{s},\boldsymbol{u}\Big]+ML^{(P_{S})}\mathbb{P}\Big[i_{n}^{(P_{S})}(\overline{\boldsymbol{U}},\boldsymbol{Y})>\gamma^{(P_{S})}\,\Big|\,\boldsymbol{s},\boldsymbol{u}\Big].\label{eq:DPC_ThresholdBound2}
\end{equation}
The analysis of the second term in \eqref{eq:DPC_ThresholdBound2}
is simplified by the following lemma.
\begin{lem}
\label{lem:DPC_Property1} Fix $P_{S}\in\tilde{\mathcal{P}}_{n}$
and $(\boldsymbol{s},\boldsymbol{u})$ such that $f_{\boldsymbol{S}\boldsymbol{U}}^{(P_{S})}(\boldsymbol{s},\boldsymbol{u})>0$,
and define the random variables 
\begin{align}
(\boldsymbol{X}^{\prime}\,|\,\boldsymbol{s},\boldsymbol{u}) & \sim\frac{\delta\big(\|\boldsymbol{x}^{\prime}\|-\|\boldsymbol{u}+(1-\alpha)\boldsymbol{s}\|\big)}{S_{n}\big(\|\boldsymbol{u}+(1-\alpha)\boldsymbol{s}\|\big)}\label{eq:DPC_X'}\\
\boldsymbol{Y}^{\prime} & =\boldsymbol{X}^{\prime}+\boldsymbol{Z},
\end{align}
where $S_{n}$ is defined in \eqref{eq:DPC_Sn}, and $\boldsymbol{Z}$
is the additive noise in \eqref{eq:DPC_Channel}. For $\overline{\boldsymbol{U}}\sim f_{\boldsymbol{U}}^{(P_{S})}$
independent of $(\boldsymbol{S},\boldsymbol{U},\boldsymbol{Y},\boldsymbol{X}^{\prime},\boldsymbol{Y}^{\prime})$,
we have
\begin{equation}
\mathbb{P}\Big[i_{n}^{(P_{S})}(\overline{\boldsymbol{U}},\boldsymbol{Y})>\gamma^{(P_{S})}\,\Big|\,\boldsymbol{s},\boldsymbol{u}\Big]=\mathbb{P}\Big[i_{n}^{(P_{S})}(\overline{\boldsymbol{U}},\boldsymbol{Y}^{\prime})>\gamma^{(P_{S})}\,\Big|\,\boldsymbol{s},\boldsymbol{u}\Big].\label{eq:DPC_SwapDistr}
\end{equation}
Furthermore, letting $f_{\boldsymbol{Y}^{\prime}|\boldsymbol{S}\boldsymbol{U}}^{(P_{S})}$
denote the density of $\boldsymbol{Y}^{\prime}$ given $(\boldsymbol{s},\boldsymbol{u})$,
there exists $\epsilon>0$ such that
\begin{align}
\mathbb{P}\Big[\big|\|\boldsymbol{Y}^{\prime}\|^{2}-n(P+P_{S}+1)\big|>n\epsilon\,\Big|\,\boldsymbol{s},\boldsymbol{u}\Big] & =O\big(e^{-\psi n}\big)\label{eq:DPC_Y'tail}\\
\min_{\boldsymbol{y}^{\prime}\,:\,\big|\|\boldsymbol{y}^{\prime}\|^{2}-n(P+P_{S}+1)\big|\le n\epsilon}\frac{f_{\boldsymbol{Y}^{\prime}|\boldsymbol{S}\boldsymbol{U}}^{(P_{S})}(\boldsymbol{y}^{\prime}|\boldsymbol{s},\boldsymbol{u})}{\prod_{i=1}^{n}f_{Y}^{(P_{S})}(y_{i}^{\prime})} & \le K_{2}\label{eq:DPC_ChgMeasure}
\end{align}
for some constants $\psi>0$ and $K_{2}$ not depending on $P_{S}$,
where $f_{Y}^{(P_{S})}$ is defined in \eqref{eq:DPC_fY}.\end{lem}
\begin{IEEEproof}
See Appendix \ref{sub:DPC_PF_PROPS}.
\end{IEEEproof}
Using Lemma \ref{lem:DPC_Property1}, we can bound the second probability
in \eqref{eq:DPC_ThresholdBound2} as follows:
\begin{align}
 & \mathbb{P}\Big[i_{n}^{(P_{S})}(\overline{\boldsymbol{U}},\boldsymbol{Y})>\gamma^{(P_{S})}\,\big|\,\boldsymbol{s},\boldsymbol{u}\Big]\nonumber \\
 & \qquad=\int_{\mathbb{R}^{n}}\int_{\mathbb{R}^{n}}f_{\boldsymbol{U}}^{(P_{S})}(\overline{\boldsymbol{u}})f_{\boldsymbol{Y}^{\prime}|\boldsymbol{S}\boldsymbol{U}}^{(P_{S})}(\boldsymbol{y}^{\prime}|\boldsymbol{s},\boldsymbol{u})\openone\big\{ i_{n}^{(P_{S})}(\overline{\boldsymbol{u}},\boldsymbol{y}^{\prime})>\gamma^{(P_{S})}\big\} d\overline{\boldsymbol{u}}d\boldsymbol{y}^{\prime}\label{eq:DPC_ThreshAnalysis1a}\\
 & \qquad=\int_{\big|\|\boldsymbol{y}^{\prime}\|^{2}-n(P+P_{S}+1)\big|\le n\epsilon}\int_{\mathbb{R}^{n}}f_{\boldsymbol{U}}^{(P_{S})}(\overline{\boldsymbol{u}})f_{\boldsymbol{Y}^{\prime}|\boldsymbol{S}\boldsymbol{U}}^{(P_{S})}(\boldsymbol{y}^{\prime}|\boldsymbol{s},\boldsymbol{u})\openone\big\{ i_{n}^{(P_{S})}(\overline{\boldsymbol{u}},\boldsymbol{y}^{\prime})>\gamma^{(P_{S})}\big\} d\overline{\boldsymbol{u}}d\boldsymbol{y}^{\prime}+O\big(e^{-\psi n}\big)\label{eq:DPC_ThreshAnalysis1b}\\
 & \qquad\le K_{2}e^{-\gamma^{(P_{S})}}+e^{-\psi n},\label{eq:DPC_ThreshAnalysis1c}
\end{align}
where \eqref{eq:DPC_ThreshAnalysis1b} follows from \eqref{eq:DPC_Y'tail},
and \eqref{eq:DPC_ThreshAnalysis1c} follows by upper bounding $f_{\boldsymbol{Y}^{\prime}|\boldsymbol{S}\boldsymbol{U}}^{(P_{S})}$
using \eqref{eq:DPC_ChgMeasure} and following the steps in \eqref{eq:GP_ThreshAnalysis1a}--\eqref{eq:GP_ThreshAnalysis1d}.

We choose
\begin{equation}
\gamma^{(P_{S})}=\log M+nI^{(P_{S})}(U;S)+\log n,\label{eq:DPC_Gamma0}
\end{equation}
which, when combined with \eqref{eq:DPC_RL}, yields 
\begin{equation}
\gamma^{(P_{S})}=\log M+nI^{(P_{S})}(U;S)+K_{3}\log n\label{eq:DPC_Gamma}
\end{equation}
with $K_{3}\triangleq K_{1}+1$. Combining \eqref{eq:DPC_ThresholdBound2}
and \eqref{eq:DPC_ThreshAnalysis1c} with this choice of $\gamma^{(P_{S})}$,
we conclude that
\begin{equation}
\mathbb{P}\big[\mathcal{E}_{2}\,|\, P_{S},\mathcal{E}_{1}^{c}\big]\le\mathbb{P}\Big[i_{n}^{(P_{S})}(\boldsymbol{u},\boldsymbol{Y})\le\log M+nI^{(P_{S})}(U;S)+K_{3}\log n\,\Big|\,\boldsymbol{s},\boldsymbol{u}\Big]+O\Big(\frac{1}{n}\Big)\label{eq:DPC_E2final}
\end{equation}
for some $(\boldsymbol{s},\boldsymbol{u})$ such that $f_{\boldsymbol{S}\boldsymbol{U}}^{(P_{S})}(\boldsymbol{s},\boldsymbol{u})>0$.

\subsubsection{Application of the Berry-Esseen Theorem}

The moments associated with $i_{n}^{(P_{S})}(\boldsymbol{u},\boldsymbol{Y})$
required to apply the Berry-Esseen theorem are characterized in the
following lemma.
\begin{lem}
\label{lem:DPC_Property2} Fix $P_{S}\in\tilde{\mathcal{P}}_{n}$
and $(\boldsymbol{s},\boldsymbol{u})$ such that $f_{\boldsymbol{S}\boldsymbol{U}}^{(P_{S})}(\boldsymbol{s},\boldsymbol{u})>0$.
We have
\begin{align}
\mathbb{E}\big[i_{n}^{(P_{S})}(\boldsymbol{u},\boldsymbol{Y})\,|\,\boldsymbol{s},\boldsymbol{u}\big] & =nI^{(P_{S})}(U;Y)+O(1),\label{eq:DPC_CondAvg}
\end{align}
and for $\alpha=\frac{P}{1+P}$ we have
\begin{equation}
\mathrm{Var}\big[i_{n}^{(P_{S})}(\boldsymbol{u},\boldsymbol{Y})\,|\,\boldsymbol{s},\boldsymbol{u}\big]=nV+O(1),\label{eq:DPC_CondVar}
\end{equation}
where $V$ is defined in \eqref{eq:DPC_V}. Furthermore, there exists
a pair $(\boldsymbol{s}^{\prime},\boldsymbol{u}^{\prime})$ such that
$i_{n}^{(P_{S})}(\boldsymbol{U},\boldsymbol{Y})$ has the same distribution
whether conditioned on \textup{$(\boldsymbol{S},\boldsymbol{U})=(\boldsymbol{s}^{\prime},\boldsymbol{u}^{\prime})$
or} $(\boldsymbol{S},\boldsymbol{U})=(\boldsymbol{s},\boldsymbol{u})$,
and such that 
\begin{equation}
\sum_{i=1}^{n}\mathbb{E}\bigg[\Big|i^{(P_{S})}(u_{i}^{\prime},Y_{i})-\mathbb{E}\big[i^{(P_{S})}(u_{i}^{\prime},Y_{i})\big]\Big|^{3}\,\Big|\, s_{i}^{\prime},u_{i}^{\prime}\bigg]=O(n).\label{eq:DPC_CondThird}
\end{equation}
The remainder terms in \eqref{eq:DPC_CondAvg}--\eqref{eq:DPC_CondThird}
are uniform in $P_{S}\in\tilde{\mathcal{P}}_{n}$.\end{lem}
\begin{IEEEproof}
See Appendix \ref{sub:DPC_PF_PROPS}.
\end{IEEEproof}
Combining \eqref{eq:DPC_E1decay} and \eqref{eq:DPC_E2final}, we
have some $(\boldsymbol{s},\boldsymbol{u})$ that
\begin{align}
\mathbb{P}\big[\mathcal{E}_{1}\cup\mathcal{E}_{2}\,|\, P_{S}\big] & \le\mathbb{P}\Big[i_{n}^{(P_{S})}(\boldsymbol{u},\boldsymbol{Y})\le\log M+nI^{(P_{S})}(U;S)+K_{3}\log n\,\Big|\,\boldsymbol{s},\boldsymbol{u}\Big]+O\Big(\frac{1}{n}\Big)\\
 & \le\mathbb{P}\Big[i_{n}^{(P_{S})}(\boldsymbol{u},\boldsymbol{Y})-\mathbb{E}\big[i_{n}^{(P_{S})}(\boldsymbol{u},\boldsymbol{Y})\big]\le\log M-nI(P_{S})+K_{3}\log n+K_{4}\,\Big|\,\boldsymbol{s},\boldsymbol{u}\Big]+O\Big(\frac{1}{n}\Big),\label{eq:DPC_App2}
\end{align}
where \eqref{eq:DPC_App2} follows from \eqref{eq:DPC_CondAvg} and
by defining 
\begin{equation}
I(P_{S})\triangleq I^{(P_{S})}(U;Y)-I^{(P_{S})}(U;S).
\end{equation}
The constant $K_{4}$ in \eqref{eq:DPC_App2} represents the uniform
$O(1)$ term in \eqref{eq:DPC_CondAvg}.

The mutual informations $I^{(P_{S})}(U;Y)$ and $I^{(P_{S})}(U;S)$
are given in \eqref{eq:DPC_IUY_Ps}--\eqref{eq:DPC_IUS_Ps}, and similarly
to \eqref{eq:DPC_C}, setting $\alpha=\frac{P}{1+P}$ yields $I(P_{S})=C$
for all $P_{S}$. Thus, applying the Berry-Esseen theorem \cite[Sec. XVI.5]{Feller}
to \eqref{eq:DPC_App2} (after replacing $(\boldsymbol{s},\boldsymbol{u})$
by $(\boldsymbol{s}^{\prime},\boldsymbol{u}^{\prime})$ given in Lemma
\ref{lem:DPC_Property2} if necessary), we obtain for all $P_{S}\in\tilde{\mathcal{P}}_{n}$
that
\begin{equation}
\mathbb{P}\big[\mathcal{E}_{1}\cup\mathcal{E}_{2}\,|\, P_{S}\big]\le\mathsf{Q}\bigg(\frac{\log M-nC+K_{3}\log n+K_{4}}{\sqrt{nV+K_{5}}}\bigg)+O\Big(\frac{1}{\sqrt{n}}\Big)\label{eq:DPC_App4}
\end{equation}
for some constant $K_{5}$. Substituting \eqref{eq:DPC_App4} into
\eqref{eq:DPC_OverallError} yields
\begin{equation}
\overline{p}_{e}\le\mathsf{Q}\bigg(\frac{\log M-nC+K_{3}\log n+K_{4}}{\sqrt{nV+K_{5}}}\bigg)+O\Big(\frac{\log n}{\sqrt{n}}\Big),
\end{equation}
and the proof of Theorem \ref{thm:DPC_MainResult} is concluded by
inverting the relationship between the error probability and the number
of messages.

\section{Concluding Remarks \label{sec:GP_CONCLUSION}}

We have presented an achievable second-order coding rate for the discrete
memoryless Gel'fand-Pinsker channel, and a conclusive characterization
of the second-order asymptotics for dirty paper coding. Possible areas
of further research include non-asymptotic bounds and their comparison
to the normal approximations obtained by omitting the higher-order
terms in \eqref{eq:GP_MainResult} and \eqref{eq:DPC_MainResult},
and second-order converse results for the discrete case. For the latter,
the techniques used in proving the strong converse \cite{GPStrongConv1,GPStrongConv2}
may prove useful.

The assumption $\mathbb{E}\big[\mathrm{Var}[i(U,Y)\,|\, S,U]\big]>0$
in Theorem \ref{eq:GP_MainResult} is analogous to similar assumptions
of positive dispersions in other settings (e.g. see \cite{TanState,DispCompound}).
When $\mathbb{E}\big[\mathrm{Var}[i(U,Y)\,|\, S,U]\big]=0$, the analysis
in Section \ref{sec:GP_PROOFS} remains valid until \eqref{eq:GP_ThreshAnalysis4},
but there are several difficulties in generalizing the subsequent
analysis. First, it does not necessarily follow that that $\mathbb{E}\big[\mathrm{Var}[i^{(P_{S})}(U,Y)\,|\, S,U]\big]=0$
under $P_{SUY}^{(P_{S})}$ (see \eqref{eq:GP_PsDistr}) for all $P_{S}\in\tilde{\mathcal{P}}_{n}$,
and hence we may still need to consider variances of up to $O\big(\sqrt{\frac{\log n}{n}}\big)$.
Second, the behavior of the probability in \eqref{eq:GP_ThreshAnalysis4}
varies depending on whether $I(\pi)>I(P_{S})$ or $I(\pi)<I(P_{S})$,
both of which can occur with differences of up to $O\big(\sqrt{\frac{\log n}{n}}\big)$.
Finally, \cite[Lemma 18]{TanState} (which is used in the proof of
Lemma \ref{lem:GP_Lem2}) is based on properties of the $\mathsf{Q}$-function,
and thus may be difficult to extend if an alternative bound (e.g.
Chebyshev's inequality) is used in place of the Berry-Esseen theorem
following \eqref{eq:GP_ThreshAnalysis4}.

A by-product of our analysis for the Gaussian case (dirty paper coding)
is an alternative viewpoint as to why similar performance is achieved
for Gaussian or non-Gaussian state sequences: By using a small fraction
of the block to send a quantized version of the state power, we can
make the decoder aware that the sequence lies within a thin spherical
shell. Since all sequences within that shell are essentially equally
difficult to handle, the precise statistics of the state sequence
are not important.

\appendix

\subsection{\label{sub:GP_DecSIPf}Proof of Corollary \ref{cor:GP_DecStateInfo}}

We apply Theorem \ref{thm:GP_MainResult} with $(Y,S)$ in place of
$Y$. The capacity-achieving parameters are $\mathcal{U}=\mathcal{X}$,
$\phi(u,s)=u$ (i.e. $x=u$) and $Q_{U|S}(\cdot|s)=Q(\cdot|s)$, where
$Q(\cdot|s)$ achieves the capacity-dispersion pair $(C_{s},V_{s})$
for the channel $W(\cdot|\cdot,s)$. To avoid ambiguity, we denote
the resulting information densities in \eqref{eq:GP_ius}--\eqref{eq:GP_iuy}
by $i_{1}(x,s)$ and $i_{2}(x,y,s)$ respectively. Defining
\begin{equation}
P_{SXY}(s,x,y)\triangleq\pi(s)Q(x|s)W(y|x,s),
\end{equation}
we have
\begin{align}
i_{1}(x,s) & =\log\frac{P_{SX}(s,x)}{P_{S}(s)P_{X}(x)}\\
i_{2}(x,y,s) & =\log\frac{P_{SXY}(s,x,y)}{P_{X}(x)P_{SY}(s,y)}.
\end{align}
It follows that
\begin{align}
i_{2}(x,y,s)-i_{1}(x,s) & =\log\frac{P_{SXY}(s,x,y)P_{S}(s)}{P_{SY}(s,y)P_{SX}(s,x)}\\
 & =\log\frac{P_{XY|S}(x,y|s)}{P_{Y|S}(y|s)P_{X|S}(x|s)}\\
 & \triangleq i_{3}(x,y,s).
\end{align}
We observe that $i_{3}(\cdot,\cdot,s)$ is the information density
associated with $W(\cdot|\cdot,s)$, and thus has mean $C_{s}$ and
variance $V_{s}$ \cite{Finite}. It follows that $V$ in \eqref{eq:GP_V2}
can be written as 
\begin{align}
V & =\mathrm{Var}[i_{3}(X,Y,S)]\\
 & =\mathbb{E}\big[\mathrm{Var}[i_{3}(X,Y,S)\,|\, S]\big]+\mathrm{Var}\big[\mathbb{E}[i_{3}(X,Y,S)\,|\, S]\big]\label{eq:GP_CorStep8}\\
 & =\mathbb{E}[V_{S}]+\mathrm{Var}[C_{S}],
\end{align}
where \eqref{eq:GP_CorStep8} follows from the law of total variance.
We similarly have $C=\mathbb{E}[i_{3}(X,Y,S)]=\mathbb{E}[C_{S}]$,
thus completing the proof.

\subsection{Continuous Differentiability and Taylor Expansions \label{sub:GP_TAYLOR_EXP}}

In this section, we study the differentiability properties of $I(P_{S})$
(see \eqref{eq:GP_IPS}) and $V(P_{S})$ (see \eqref{eq:GP_Var2}),
and prove the Taylor expansion given \eqref{eq:GP_Taylor}.

\subsubsection{Derivatives of $I(P_{S})$ }

Writing
\begin{equation}
I^{(P_{S})}(U;S)=\sum_{s,u}P_{S}(s)Q_{U|S}(u|s)\log\frac{Q_{U|S}(u|s)}{\sum_{\overline{s}}P_{S}(\overline{s})Q_{U|S}(u|\overline{s})},\label{eq:GP_IUSfull}
\end{equation}
we obtain
\begin{align}
\frac{\partial I^{(P_{S})}(U;S)}{\partial P_{S}(s^{\prime})} & =\sum_{s\ne s^{\prime},u}P_{S}(s)Q_{U|S}(u|s)\frac{\partial}{\partial P_{S}(s^{\prime})}\bigg(-\log\sum_{\overline{s}}P_{S}(\overline{s})Q_{U|S}(u|\overline{s})\bigg)\nonumber \\
 & \qquad\qquad\qquad\qquad+\frac{\partial}{\partial P_{S}(s^{\prime})}\sum_{u}P_{S}(s^{\prime})Q_{U|S}(u|s^{\prime})\log\frac{Q_{U|S}(u|s^{\prime})}{\sum_{\overline{s}}P_{S}(\overline{s})Q_{U|S}(u|\overline{s})}\label{eq:GP_App1}\\
 & =-\sum_{s,u}P_{S}(s)Q_{U|S}(u|s)\frac{Q_{U|S}(u|s^{\prime})}{\sum_{\overline{s}}P_{S}(\overline{s})Q_{U|S}(u|\overline{s})}+\sum_{u}Q_{U|S}(u|s^{\prime})\log\frac{Q_{U|S}(u|s^{\prime})}{\sum_{\overline{s}}P_{S}(\overline{s})Q_{U|S}(u|\overline{s})}\\
 & =-\sum_{s,u}P_{S|U}(s|u)Q_{U|S}(u|s^{\prime})+\sum_{u}Q_{U|S}(u|s^{\prime})\log\frac{Q_{U|S}(u|s^{\prime})}{\sum_{\overline{s}}P_{S}(\overline{s})Q_{U|S}(u|\overline{s})}\label{eq:GP_App3}\\
 & =-1+\sum_{u}Q_{U|S}(u|s^{\prime})\log\frac{Q_{U|S}(u|s^{\prime})}{\sum_{\overline{s}}P_{S}(\overline{s})Q_{U|S}(u|\overline{s})},\label{eq:GP_App4}
\end{align}
where \eqref{eq:GP_App3} follows by writing $P_{S|U}(s|u)=\frac{P_{S}(s)Q_{U|S}(u|s)}{P_{U}(u)}$. 

The derivative of $I^{(P_{S})}(U;Y)$ is computed similarly. We have
\begin{align}
I^{(P_{S})}(U;Y) & =\sum_{s,u,y}P_{S}(s)P_{UY|S}(u,y|s)\log\frac{P_{UY}^{(P_{S})}(u,y)}{P_{U}^{(P_{S})}(u)P_{Y}^{(P_{S})}(y)}\label{eq:GP_IUYfull0}\\
 & =\sum_{s,u,y}P_{S}(s)P_{UY|S}(u,y|s)\Big(\log P_{UY}^{(P_{S})}(u,y)-\log P_{U}^{(P_{S})}(u)-\log P_{Y}^{(P_{S})}(y)\Big),\label{eq:GP_IUYfull}
\end{align}
where $P_{UY|S}(u,y|s)=Q_{U|S}(u|s)W(y|\phi(u,s),s)$ does not depend
on $P_{S}$. We can write
\begin{equation}
P_{UY}^{(P_{S})}(u,y)=\sum_{s}P_{S}(s)P_{UY|S}(u,y|s)\label{eq:GP_PUY}
\end{equation}
and similarly for $P_{U}$ and $P_{Y}$, yielding the derivatives
\begin{align}
\frac{\partial P_{UY}^{(P_{S})}(u,y)}{\partial P_{S}(s^{\prime})} & =P_{UY|S}(u,y|s^{\prime})\\
\frac{\partial P_{U}^{(P_{S})}(u)}{\partial P_{S}(s)} & =P_{U|S}(u|s^{\prime})\\
\frac{\partial P_{Y}^{(P_{S})}(y)}{\partial P_{S}(s^{\prime})} & =P_{Y|S}(y|s^{\prime}).
\end{align}
It follows using the same arguments as \eqref{eq:GP_App1}--\eqref{eq:GP_App4}
that 
\begin{align}
\frac{\partial I^{(P_{S})}(U;Y)}{\partial P_{S}(s^{\prime})} & =\sum_{s,u,y}P_{S}(s)P_{UY|S}(u,y|s)\bigg(\frac{P_{UY|S}(u,y|s^{\prime})}{P_{UY}^{(P_{S})}(u,y)}-\frac{P_{U|S}(u|s^{\prime})}{P_{U}^{(P_{S})}(u)}-\frac{P_{Y|S}(y|s^{\prime})}{P_{Y}^{(P_{S})}(y)}\bigg)\nonumber \\
 & \qquad\qquad\qquad\qquad\qquad\qquad\qquad+\sum_{u,y}P_{UY|S}(u,y|s^{\prime})\log\frac{P_{UY}^{(P_{S})}(u,y)}{P_{U}^{(P_{S})}(u)P_{Y}^{(P_{S})}(y)}\\
 & =-1+\sum_{u,y}P_{UY|S}(u,y|s^{\prime})\log\frac{P_{UY}^{(P_{S})}(u,y)}{P_{U}^{(P_{S})}(u)P_{Y}^{(P_{S})}(y)}.\label{eq:GP_App11}
\end{align}
Differentiating \eqref{eq:GP_App4} and \eqref{eq:GP_App11} a second
time, we obtain
\begin{align}
\frac{\partial^{2}I^{(P_{S})}(U;S)}{\partial P_{S}(s^{\prime})\partial P_{S}(s^{\prime\prime})} & =-\sum_{u}Q_{U|S}(u|s^{\prime})\frac{Q_{U|S}(u|s^{\prime\prime})}{\sum_{\overline{s}}P_{S}(\overline{s})Q_{U|S}(u|\overline{s})}\label{eq:GP_App12}\\
\frac{\partial^{2}I^{(P_{S})}(U;Y)}{\partial P_{S}(s^{\prime})\partial P_{S}(s^{\prime\prime})} & =\sum_{u,y}P_{UY|S}(u,y|s^{\prime})\bigg(\frac{P_{UY|S}(u,y|s^{\prime\prime})}{P_{UY}^{(P_{S})}(u,y)}-\frac{P_{U|S}(u|s^{\prime\prime})}{P_{U}^{(P_{S})}(u)}-\frac{P_{Y|S}(y|s^{\prime\prime})}{P_{Y}^{(P_{S})}(y)}\bigg).\label{eq:GP_App13}
\end{align}

\subsubsection{Continuous Differentiability }

Using \eqref{eq:GP_PUY}, we observe the derivatives in \eqref{eq:GP_App4}
and \eqref{eq:GP_App11}--\eqref{eq:GP_App13} are continuous in $P_{S}$
other than a possible divergence as $\min_{s}P_{S}(s)\to0$. Assuming
without loss of generality that $\min_{s}\pi(s)>0$, it follows that
$I(P_{S})$ is twice continuously differentiable within $\tilde{\mathcal{P}}_{n}$
(see \eqref{eq:GP_SetPtilde}) for sufficiently large $n$.

For $V(P_{S})$ (see \eqref{eq:GP_Var2}) we only require that the
first derivatives are continuous within $\tilde{\mathcal{P}}_{n}$.
This can be proved by writing $V(P_{S})$ in the form
\begin{align}
V(P_{S}) & =\sum_{s,u}P_{S}(s)Q_{U|S}(u|s)V^{(P_{S})}(u,s),
\end{align}
where
\begin{equation}
V^{(P_{S})}(u,s)\triangleq\sum_{y}W(y|\phi(u,s),s)\bigg(\log\frac{P_{UY}^{(P_{S})}(u,y)}{P_{U}^{(P_{S})}(u)P_{Y}^{(P_{S})}(y)}\bigg)^{2}-\bigg(\sum_{y}W(y|\phi(u,s),s)\log\frac{P_{UY}^{(P_{S})}(u,y)}{P_{U}^{(P_{S})}(u)P_{Y}^{(P_{S})}(y)}\bigg)^{2}.
\end{equation}
The subsequent evaluation of the partial derivatives is cumbersome
and similar to the analysis following \eqref{eq:GP_IUYfull0}, and
is thus omitted.

\subsubsection{Taylor Expansion of $I(P_{S})$}

The first-order Taylor approximation of $I(P_{S})=I^{(P_{S})}(U;Y)-I^{(P_{S})}(U;S)$
at $P_{S}=\pi$ is given by
\begin{align}
I(P_{S}) & =I(\pi)+\sum_{s}\big(P_{S}(s)-\pi(s)\big)\frac{\partial I(P_{S})}{\partial P_{S}(s)}\bigg|_{P_{S}=\pi}+\Delta(P_{S}),\label{eq:GP_App14}
\end{align}
where $\Delta(P_{S})$ is the remainder term. From \eqref{eq:GP_App4}
and \eqref{eq:GP_App11}, we see that $\sum_{s}\pi(s)\frac{\partial I(P_{S})}{\partial P_{S}(s)}\Big|_{P_{S}=\pi}=I(\pi)$,
and hence the right-hand side of \eqref{eq:GP_App14} equals $\sum_{s}P_{S}(s)\frac{\partial I(P_{S})}{\partial P_{S}(s)}\Big|_{P_{S}=\pi}+\Delta(P_{S})$,
which in turn equals $\tilde{I}(P_{S})+\Delta(P_{S})$ (see \eqref{eq:GP_Itilde}).
The remainder term satisfies \eqref{eq:GP_DeltaProperty} since $I(P_{S})$
is twice continuously differentiable within $\tilde{\mathcal{P}}_{n}$,
and since the $\ell_{2}$-norm and $\ell_{\infty}$-norm coincide
to within a constant factor.

\subsection{Necessary Conditions for the Optimal Input Distribution \label{sub:GP_NECC_CONDS}}

Here we study the necessary Karush-Kuhn-Tucker (KKT) conditions \cite[Sec. 5.5.3]{Convex}
for $Q_{U|S}$ to maximize the objective in \eqref{eq:GP_Capacity}
when $\mathcal{U}$ and $\phi(\cdot,\cdot)$ are fixed. Introducing
the Lagrange multiplier $\lambda(s)$ corresponding to the constraint
$\sum_{u}Q_{U|S}(u|s)=1$, we see that any optimal $Q_{U|S}$ must
satisfy
\begin{equation}
\frac{\partial}{\partial Q_{U|S}(u^{\prime}|s^{\prime})}\Big(I(U;Y)-I(U;S)\Big)=\lambda(s^{\prime})\label{eq:GP_KKT1}
\end{equation}
for all $(s^{\prime},u^{\prime})$ such that $Q_{U|S}(u^{\prime}|s^{\prime})>0$;
we assume without loss of generality that $\min_{s}\pi(s)>0$.

Using \eqref{eq:GP_IUSfull} with $P_{S}=\pi$, and writing the logarithm
as a difference of two logarithms, we obtain
\begin{align}
\frac{\partial I(U;S)}{\partial Q_{U|S}(u^{\prime}|s^{\prime})} & =\pi(s^{\prime})\log Q_{U|S}(u^{\prime}|s^{\prime})+\pi(s^{\prime})-\pi(s^{\prime})\log\sum_{\overline{s}}\pi(\overline{s})Q_{U|S}(u^{\prime}|\overline{s})\nonumber \\
 & \qquad\qquad\qquad\qquad\qquad\qquad-\sum_{s}\pi(s)Q_{U|S}(u^{\prime}|s)\frac{\pi(s^{\prime})}{\sum_{\overline{s}}\pi(\overline{s})Q_{U|S}(u^{\prime}|\overline{s})}\label{eq:GP_KKT3}\\
 & =\pi(s^{\prime})\log\frac{Q_{U|S}(u^{\prime}|s^{\prime})}{\sum_{\overline{s}}\pi(\overline{s})Q_{U|S}(u^{\prime}|\overline{s})},\label{eq:GP_KKT4}
\end{align}
where \eqref{eq:GP_KKT4} follows by applying Bayes' rule to the last
term in \eqref{eq:GP_KKT3}. To evaluate the partial derivatives of
$I(U;Y)$, we write \eqref{eq:GP_IUYfull} (with $P_{S}=\pi$) as
\[
I(U;Y)=\sum_{s,u,y}\pi(s)Q_{U|S}(u|s)W(y|\phi(u,s),s)\Big(\log P_{UY}(u,y)-\log P_{U}(u)-\log P_{Y}(y)\Big).
\]
We have $P_{UY}(u,y)=\sum_{s}\pi(s)Q_{U|S}(u|s)W(y|\phi(u,s),s)$
(and similarly for $P_{U}$ and $P_{Y}$), yielding the derivatives
\begin{align}
\frac{\partial P_{UY}(u,y)}{\partial Q_{U|S}(u^{\prime}|s^{\prime})} & =\begin{cases}
\pi(s^{\prime})W(y|\phi(u^{\prime},s^{\prime}),s^{\prime}) & u=u^{\prime}\\
0 & \mathrm{otherwise}
\end{cases}\\
\frac{\partial P_{U}(u)}{\partial Q_{U|S}(u^{\prime}|s^{\prime})} & =\begin{cases}
\pi(s^{\prime}) & u=u^{\prime}\\
0 & \mathrm{otherwise}
\end{cases}\\
\frac{\partial P_{Y}(y)}{\partial Q_{U|S}(u^{\prime}|s^{\prime})} & =\pi(s^{\prime})W(y|\phi(u^{\prime},s^{\prime}),s^{\prime}).
\end{align}
It follows using a similar argument to \eqref{eq:GP_KKT3}--\eqref{eq:GP_KKT4}
that
\begin{equation}
\frac{\partial I(U;Y)}{\partial Q_{U|S}(u^{\prime}|s^{\prime})}=\pi(s^{\prime})\Bigg(\sum_{y}W(y|\phi(u^{\prime},s^{\prime}),s^{\prime})\log\frac{P_{UY}(u^{\prime},y)}{P_{U}(u^{\prime})P_{Y}(y)}-1\Bigg).\label{eq:GP_KKT8}
\end{equation}
Combining \eqref{eq:GP_KKT1}, \eqref{eq:GP_KKT4} and \eqref{eq:GP_KKT8},
we see that for any optimal $Q_{U|S}$ and any $s\in\mathcal{S}$,
the quantity
\begin{equation}
\sum_{y}W(y|\phi(u,s),s)\log\frac{P_{UY}(u,y)}{P_{U}(u)P_{Y}(y)}-\log\frac{Q_{U|S}(u|s)}{\sum_{\overline{s}}\pi(\overline{s})Q_{U|S}(u|\overline{s})}\label{eq:GP_KKT9}
\end{equation}
is the same for all $u$ such that $Q_{U|S}(u|s)>0$. Note that \eqref{eq:GP_KKT9}
can be written more compactly as $\mathbb{E}[i(u,Y)-i(u,s)\,|\, s,u]$;
see \eqref{eq:GP_ius}--\eqref{eq:GP_iuy}.

\subsection{Proofs of Steps Involving Taylor Expansions \label{sub:GP_PROOFS_STEPS}}

In this section, we make use of the fact that
\begin{equation}
\min_{P_{S}\in\tilde{\mathcal{P}}_{n}}V(P_{S})\ge V_{\mathrm{min}}\label{eq:GP_Vmin}
\end{equation}
for some $V_{\mathrm{min}}>0$ and sufficiently large $n$, which
follows from the definition of $V(P_{S})$ in \eqref{eq:GP_Var2},
the assumption of Theorem \ref{thm:GP_MainResult}, and the fact that
$P_{S}\to\pi$ within $\tilde{\mathcal{P}}_{n}$ (see \eqref{eq:GP_SetPtilde}).

\subsubsection{Proof of \eqref{eq:GP_AvgOverS2}}

We first eliminate $K_{5}$ from \eqref{eq:GP_AvgOverS} by writing
\begin{equation}
\sum_{P_{S}\in\tilde{\mathcal{P}}_{n}}\mathbb{P}[P_{S}]\mathsf{Q}\bigg(\frac{\beta_{n}+nI(P_{S})-nI(\pi)+K_{5}}{\sqrt{nV(P_{S})+K_{6}}}\bigg)=\sum_{P_{S}\in\tilde{\mathcal{P}}_{n}}\mathbb{P}[P_{S}]\mathsf{Q}\bigg(\frac{\beta_{n}+nI(P_{S})-nI(\pi)}{\sqrt{nV(P_{S})+K_{6}}}\bigg)+O\Big(\frac{1}{\sqrt{n}}\Big),
\end{equation}
which follows from \eqref{eq:GP_Vmin} and the identity $|\mathsf{Q}(z)-\mathsf{Q}(z+a)|\le\frac{|a|}{\sqrt{2\pi}}$.
It remains to eliminate $K_{6}$. The case $K_{6}\le0$ is trivial,
so we assume that $K_{6}>0$. We have
\begin{align}
\frac{1}{\sqrt{nV(P_{S})+K_{6}}} & =\frac{1}{\sqrt{nV(P_{S})}\sqrt{1+\frac{K_{6}}{nV(P_{S})}}}\\
 & \ge\frac{1}{\sqrt{nV(P_{S})}}\Big(1-\frac{K_{6}^{\prime}}{n}\Big),\label{eq:GP_UnifProof3}
\end{align}
where \eqref{eq:GP_UnifProof3} follows with $K_{6}^{\prime}=\frac{K_{6}}{2V_{\mathrm{min}}}$
using \eqref{eq:GP_Vmin} and the identity $\frac{1}{\sqrt{1+z}}\ge1-\frac{z}{2}$.
We thus obtain
\begin{align}
 & \sum_{P_{S}\in\tilde{\mathcal{P}}_{n}}\mathbb{P}[P_{S}]\mathsf{Q}\bigg(\frac{\beta_{n}+nI(P_{S})-nI(\pi)}{\sqrt{nV(P_{S})+K_{6}}}\bigg)\\
 & \quad\le\sum_{P_{S}\in\tilde{\mathcal{P}}_{n}}\mathbb{P}[P_{S}]\mathsf{Q}\bigg(\frac{\beta_{n}+nI(P_{S})-nI(\pi)}{\sqrt{nV(P_{S})}}\Big(1-\frac{K_{6}^{\prime}}{n}\Big)\bigg)\\
 & \quad\le\sum_{P_{S}\in\tilde{\mathcal{P}}_{n}}\mathbb{P}[P_{S}]\mathsf{Q}\bigg(\frac{\beta_{n}+nI(P_{S})-nI(\pi)}{\sqrt{nV(P_{S})}}\bigg)+\frac{K_{6}^{\prime}}{\sqrt{2\pi}n}\sum_{P_{S}\in\tilde{\mathcal{P}}_{n}}\mathbb{P}\big[P_{S}\big]\frac{|\beta_{n}+nI(P_{S})-nI(\pi)|}{\sqrt{nV(P_{S})}}\label{eq:GP_UnifProof6}\\
 & \quad=\sum_{P_{S}\in\tilde{\mathcal{P}}_{n}}\mathbb{P}[P_{S}]\mathsf{Q}\bigg(\frac{\beta_{n}+nI(P_{S})-nI(\pi)}{\sqrt{nV(P_{S})}}\bigg)+O\Big(\frac{1}{\sqrt{n}}\Big),\label{eq:GP_UnifProof7}
\end{align}
where \eqref{eq:GP_UnifProof6} follows from the identity $|\mathsf{Q}(z)-\mathsf{Q}(z+a)|\le\frac{|a|}{\sqrt{2\pi}}$,
and \eqref{eq:GP_UnifProof7} follows since $\beta_{n}=O(\sqrt{n})$
by assumption.

\subsubsection{Upper Bound on \eqref{eq:GP_Lem1_2}}

Using the same argument as the one leading to \eqref{eq:GP_UnifProof6},
we have for some $K_{7}^{\prime}$ that
\begin{align}
 & \sum_{P_{S}\in\tilde{\mathcal{P}}_{n}}\mathbb{P}[P_{S}]\mathsf{Q}\Bigg(\frac{\beta_{n}+nI(P_{S})-nI(\pi)}{\sqrt{n\big(V(\pi)+K_{7}\sqrt{\frac{\log n}{n}}\big)}}\Bigg)\nonumber \\
 & \quad\le\sum_{P_{S}\in\tilde{\mathcal{P}}_{n}}\mathbb{P}[P_{S}]\mathsf{Q}\Bigg(\frac{\beta_{n}+nI(P_{S})-nI(\pi)}{\sqrt{nV(\pi)}}\Bigg)+K_{7}^{\prime}\sqrt{\frac{\log n}{n}}\sum_{P_{S}\in\tilde{\mathcal{P}}_{n}}\mathbb{P}[P_{S}]\frac{|\beta_{n}+nI(P_{S})-nI(\pi)|}{\sqrt{nV(P_{S})}}.\label{eq:GP_Lem1Step1}
\end{align}
 We now analyze the growth rate of the second term. Applying \eqref{eq:GP_Vmin},
we can upper bound this term by
\begin{equation}
\frac{K_{7}^{\prime}}{\sqrt{V_{\mathrm{min}}}}\frac{\sqrt{\log n}}{n}\sum_{P_{S}\in\tilde{\mathcal{P}}_{n}}\mathbb{P}[P_{S}]\big|\beta_{n}+nI(P_{S})-nI(\pi)\big|.
\end{equation}
Since $I(\cdot)$ is continuously differentiable (see Appendix \ref{sub:GP_TAYLOR_EXP}),
we have from \eqref{eq:GP_SetPtilde} that $\max_{P_{S}\in\tilde{\mathcal{P}}_{n}}|I(P_{S})-I(\pi)|\le K_{8}\sqrt{\frac{\log n}{n}}$
for some constant $K_{8}$. Using this observation along with $\beta_{n}=O(\sqrt{n})$,
we obtain
\begin{align}
\frac{K_{7}^{\prime}}{\sqrt{V_{\mathrm{min}}}}\frac{\sqrt{\log n}}{n}\sum_{P_{S}\in\tilde{\mathcal{P}}_{n}}\mathbb{P}[P_{S}]\big|\beta_{n}+nI(P_{S})-nI(\pi)\big| & \le\frac{K_{7}^{\prime}}{\sqrt{V_{\mathrm{min}}}}\frac{\sqrt{\log n}}{n}\sum_{P_{S}\in\tilde{\mathcal{P}}_{n}}\mathbb{P}[P_{S}]\Big(|\beta_{n}|+K_{8}\sqrt{n\log n}\Big)\\
 & =O\Big(\frac{\log n}{\sqrt{n}}\Big).\label{eq:GP_Lema1Step4}
\end{align}
Substituting \eqref{eq:GP_Lema1Step4} into \eqref{eq:GP_Lem1Step1},
we obtain the desired result.

\subsection{Proof of Lemma \ref{lem:DPC_Geometric} \label{sub:DPC_PF_GEOMETRIC}}

Recall that $\|\boldsymbol{\boldsymbol{U}}\|^{2}=n(P+\alpha^{2}P_{S})$
almost surely and $\boldsymbol{s}\in T^{n}(P_{S})$ by assumption,
and let $(\boldsymbol{s},\boldsymbol{u})$ be fixed accordingly. Writing
$\|\boldsymbol{u}-\alpha\boldsymbol{s}\|^{2}=\|\boldsymbol{u}\|^{2}-2\alpha\langle\boldsymbol{s},\boldsymbol{u}\rangle+\alpha^{2}\|\boldsymbol{s}\|^{2}$,
we have from \eqref{eq:DPC_Xcond} that $\boldsymbol{u}-\alpha\boldsymbol{s}\in\mathcal{D}_{n}$
if and only if
\begin{equation}
-n\alpha^{2}P_{S}-\delta_{x}\le-2\alpha\langle\boldsymbol{s},\boldsymbol{u}\rangle+\alpha^{2}\|\boldsymbol{s}\|^{2}\le-n\alpha^{2}P_{S},\label{eq:DPC_Geometric1}
\end{equation}
or equivalently
\begin{equation}
\frac{n\alpha P_{S}}{2}\le\langle\boldsymbol{s},\boldsymbol{u}\rangle-\frac{\alpha}{2}\|\boldsymbol{s}\|^{2}\le\frac{n\alpha P_{S}}{2}+\frac{\delta_{x}}{2\alpha}.\label{eq:DPC_Geometric2}
\end{equation}
By symmetry, the distribution of $\langle\boldsymbol{s},\boldsymbol{U}\rangle$
depends on $\boldsymbol{s}$ only through its magnitude, and we can
thus assume that $\boldsymbol{s}=(\|\boldsymbol{s}\|,0,\cdots,0)$.
In this case, the condition in \eqref{eq:DPC_Geometric2} becomes
\begin{equation}
\frac{n\alpha P_{S}}{2}\le u_{1}\|\boldsymbol{s}\|-\frac{\alpha}{2}\|\boldsymbol{s}\|^{2}\le\frac{n\alpha P_{S}}{2}+\frac{\delta_{x}}{2\alpha}.\label{eq:DPC_Geometric3}
\end{equation}
where $u_{1}$ is the first entry of $\boldsymbol{u}$. Adding $\frac{\alpha}{2}\|\boldsymbol{s}\|^{2}$
and dividing by $\|\boldsymbol{s}\|$, this becomes 
\begin{equation}
\frac{n\alpha P_{S}}{2\|\boldsymbol{s}\|}+\frac{\alpha}{2}\|\boldsymbol{s}\|\le u_{1}\le\frac{n\alpha P_{S}}{2\|\boldsymbol{s}\|}+\frac{\alpha}{2}\|\boldsymbol{s}\|+\frac{\delta_{x}}{2\alpha\|\boldsymbol{s}\|}.\label{eq:DPC_Geometric4}
\end{equation}
From \eqref{eq:DPC_TypicalSet}, there exists $P_{\mathrm{max}}<\infty$
such that $\|\boldsymbol{s}\|\le\sqrt{nP_{\mathrm{max}}}$ whenever
$P_{S}\in\tilde{\mathcal{P}}_{n}$. It follows that $\boldsymbol{u}-\alpha\boldsymbol{s}\in\mathcal{D}_{n}$
provided that
\begin{equation}
\frac{n\alpha P_{S}}{2\|\boldsymbol{s}\|}+\frac{\alpha}{2}\|\boldsymbol{s}\|\le u_{1}\le\frac{n\alpha P_{S}}{2\|\boldsymbol{s}\|}+\frac{\alpha}{2}\|\boldsymbol{s}\|+\frac{\delta_{x}}{2\alpha\sqrt{nP_{\mathrm{max}}}}.\label{eq:DPC_Geometric5}
\end{equation}
We conclude that $\mathbb{P}[\boldsymbol{U}-\alpha\boldsymbol{s}\in\mathcal{D}_{n}]$
is lower bounded by the probability of the first entry $U_{1}$ of
$\boldsymbol{U}$ falling within an interval of length $\frac{c}{\sqrt{n}}$
starting at $\frac{n\alpha P_{S}}{2\|\boldsymbol{s}\|}+\frac{\alpha}{2}\|\boldsymbol{s}\|$,
where $c\triangleq\frac{\delta_{x}}{2\alpha\sqrt{P_{\mathrm{max}}}}$.
The distribution of a given symbol in a length-$n$ random sequence
distributed uniformly on the sphere is known \cite[Eq. (4)]{SphericalDistr},
and yields
\begin{equation}
f_{U_{1}}(u_{1})=\frac{1}{\sqrt{\pi n(P+\alpha^{2}P_{S})}}\frac{\Gamma(\frac{n}{2})}{\Gamma(\frac{n-1}{2})}\bigg(1-\frac{u_{1}^{2}}{n(P+\alpha^{2}P_{S})}\bigg)^{\frac{n-3}{2}}\openone\big\{ u_{1}^{2}\le n(P+\alpha^{2}P_{S})\big\}.\label{eq:DPC_ShellDistr}
\end{equation}
This density function is decreasing in $u_{1}^{2}$, which implies
that
\begin{equation}
\mathbb{P}[\boldsymbol{U}-\alpha\boldsymbol{s}\in\mathcal{D}_{n}]\ge\frac{c}{\sqrt{n}}f_{U_{1}}\bigg(\frac{n\alpha P_{S}}{2\|\boldsymbol{s}\|}+\frac{\alpha}{2}\|\boldsymbol{s}\|+\frac{c}{\sqrt{n}}\bigg).\label{eq:DPC_Geomertic6}
\end{equation}
Furthermore, we have from \eqref{eq:DPC_SetTn} that $nP_{S}\le\|\boldsymbol{s}\|^{2}$,
and hence 
\begin{align}
\frac{n\alpha P_{S}}{2\|\boldsymbol{s}\|}+\frac{\alpha}{2}\|\boldsymbol{s}\| & \le\alpha\|\boldsymbol{s}\|\\
 & \le\alpha\sqrt{nP_{S}}+\frac{\delta_{s}}{2\sqrt{nP_{S}}},\label{eq:DPC_Geometric8}
\end{align}
where \eqref{eq:DPC_Geometric8} follows by again using \eqref{eq:DPC_SetTn},
along with the identity $\sqrt{1+\alpha}\le1+\frac{\alpha}{2}$. Thus,
the square of the argument to $f_{U_{1}}$ in \eqref{eq:DPC_Geomertic6}
is upper bounded by
\begin{align}
 & \bigg(\alpha\sqrt{nP_{S}}+\frac{\delta_{s}}{2\sqrt{nP_{S}}}+\frac{c}{\sqrt{n}}\bigg)^{2}\nonumber \\
 & \qquad=n\alpha^{2}P_{S}+2\alpha\sqrt{nP_{S}}\bigg(\frac{\delta_{s}}{2\sqrt{nP_{S}}}+\frac{c}{\sqrt{n}}\bigg)+\bigg(\frac{\delta_{s}}{2\sqrt{nP_{S}}}+\frac{c}{\sqrt{n}}\bigg)^{2}\\
 & \qquad\le n\alpha^{2}P_{S}+\alpha\delta_{s}+2\alpha\sqrt{P_{\mathrm{max}}}c+\bigg(\frac{\delta_{s}}{2\sqrt{nP_{S}}}+\frac{c}{\sqrt{n}}\bigg)^{2}\\
 & \qquad\le n\alpha^{2}P_{S}+c^{\prime},\label{eq:DPC_Geometric11}
\end{align}
where \eqref{eq:DPC_Geometric11} holds for any $c^{\prime}>\alpha\delta_{s}+2\alpha\sqrt{P_{\mathrm{max}}}c$
and sufficiently large $n$. Substituting \eqref{eq:DPC_Geometric11}
into \eqref{eq:DPC_Geomertic6} and again using the fact that $f_{U_{1}}(u_{1})$
is decreasing in $u_{1}^{2}$, we obtain 
\begin{align}
\mathbb{P}[\boldsymbol{U}-\alpha\boldsymbol{s}\in\mathcal{D}_{n}] & \ge\frac{1}{p_{0}^{\prime}(n)}\bigg(1-\frac{n\alpha^{2}P_{S}+c^{\prime}}{n(P+\alpha^{2}P_{S})}\bigg)^{\frac{n-3}{2}}\\
 & =\frac{1}{p_{0}^{\prime}(n)}\bigg(\frac{P}{P+\alpha^{2}P_{S}}\Big(1-\frac{c^{\prime}}{nP}\Big)\bigg)^{\frac{n-3}{2}}\\
 & \ge\frac{1}{p_{0}(n)}\bigg(\frac{P}{P+\alpha^{2}P_{S}}\bigg)^{\frac{n}{2}},\label{eq:DPC_Geometric10}
\end{align}
where $p_{0}^{\prime}(n)\triangleq\Big(\frac{c}{\sqrt{n}}\frac{1}{\sqrt{\pi n(P+\alpha^{2}P_{\mathrm{max}})}}\frac{\Gamma(\frac{n}{2})}{\Gamma(\frac{n-1}{2})}\Big)^{-1}$
(which grows at most polynomially fast since $\frac{\Gamma(\frac{n}{2})}{\Gamma(\frac{n-1}{2})}$
grows as $\Theta(\sqrt{n})$), and \eqref{eq:DPC_Geometric10} holds
for some polynomial $p_{0}(n)$ and sufficiently large $n$ since
$\big(1+\frac{c^{\prime}}{nP}\big)^{n/2}\to\exp\big(\frac{c^{\prime}}{2P}\big)$.
We obtain \eqref{eq:DPC_Geometric} by combining \eqref{eq:DPC_Geometric10}
with the definition of $I^{(P_{S})}(U;S)$ in \eqref{eq:DPC_IUS_Ps}.

\subsection{Proofs of Lemmas \ref{lem:DPC_Property1} and \ref{lem:DPC_Property2}
\label{sub:DPC_PF_PROPS}}

We first introduce some results which will be used in both proofs.
Recall that the information density $i^{(P_{S})}(u,y)$ is defined
with respect to the joint distribution $f_{SUY}^{(P_{S})}$ defined
in \eqref{eq:DPC_fSUY}. The corresponding covariance matrix associated
with $(S,U,Y)$ is given by
\begin{align}
\boldsymbol{V}_{SUY} & =\left[\begin{array}{ccc}
P_{S} & \alpha P_{S} & P_{S}\\
\alpha P_{S} & P+\alpha^{2}P_{S} & P+\alpha P_{S}\\
P_{S} & P+\alpha P_{S} & P+P_{S}+1
\end{array}\right].\label{eq:VSU}
\end{align}
Substituting the (Gaussian) marginal distributions $f_{Y|U}^{(P_{S})}$
and $f_{Y}^{(P_{S})}$ into \eqref{eq:DPC_iuy}, it can be shown that
\begin{align}
i^{(P_{S})}(u,y) & =\frac{1}{2}\log\frac{(P+P_{S}+1)(P+\alpha^{2}P_{S})}{PP_{S}(1-\alpha)^{2}+(P+\alpha^{2}P_{S})}\nonumber \\
 & \qquad-\frac{P+\alpha^{2}P_{S}}{2(PP_{S}(1-\alpha)^{2}+(P+\alpha^{2}P_{S}))}\bigg(y-\frac{P+\alpha P_{S}}{P+\alpha^{2}P_{S}}u\bigg)^{2}+\frac{y^{2}}{2(P+P_{S}+1)}.\label{eq:DPC_Prop5}
\end{align}
Observe that the leading term coincides with the mutual information
in \eqref{eq:DPC_IUY_Ps}.

Let $(\boldsymbol{s},\boldsymbol{u})$ be an arbitrary pair on the
support of $f_{\boldsymbol{S}\boldsymbol{U}}^{(P_{S})}$, and recall
that the definition of $f_{\boldsymbol{S}\boldsymbol{U}}^{(P_{S})}$
conditions on $\boldsymbol{S}\in T^{n}(P_{S})$ and $\mathcal{E}_{1}^{c}$.
It follows that $\|\boldsymbol{s}\|^{2}$ is bounded according to
\eqref{eq:DPC_SetTn}, and $\|\boldsymbol{x}\|^{2}=\|\boldsymbol{u}-\alpha\boldsymbol{s}\|^{2}$
is bounded according to \eqref{eq:DPC_Xcond}. It will prove useful
to show that there exists a constant $\delta_{xs}>0$ such that
\begin{equation}
n(P+P_{S})-\delta_{xs}\le\|\boldsymbol{u}+(1-\alpha)\boldsymbol{s}\|^{2}\le n(P+P_{S})+\delta_{xs}.\label{eq:DPC_Prop1}
\end{equation}
To see this, we first combine \eqref{eq:DPC_SetTn} and \eqref{eq:DPC_Geometric2}
(the latter of which was derived using only \eqref{eq:DPC_Xcond}
and the fact that $\|\boldsymbol{u}\|^{2}=n(P+\alpha^{2}P_{S})$)
to obtain
\begin{equation}
n\alpha P_{S}\le\langle\boldsymbol{s},\boldsymbol{u}\rangle\le n\alpha P_{S}+\frac{\delta_{x}}{2\alpha}+\frac{\alpha\delta_{s}}{2}.\label{eq:DPC_Prop3}
\end{equation}
Writing $\|\boldsymbol{u}+(1-\alpha)\boldsymbol{s}\|^{2}=\|\boldsymbol{u}\|^{2}+2(1-\alpha)\langle\boldsymbol{s},\boldsymbol{u}\rangle+(1-\alpha)^{2}\|\boldsymbol{s}\|^{2}$
and applying \eqref{eq:DPC_SetTn}, \eqref{eq:DPC_Prop3} and $\|\boldsymbol{u}\|^{2}=n(P+\alpha^{2}P_{S})$,
we obtain \eqref{eq:DPC_Prop1}.

\subsubsection{Proof of Lemma \ref{lem:DPC_Property1} }

To prove \eqref{eq:DPC_SwapDistr}, we will show that conditioned
on $(\boldsymbol{S},\boldsymbol{U})=(\boldsymbol{s},\boldsymbol{u})$,
the distribution of $i_{n}^{(P_{S})}(\overline{\boldsymbol{U}},\boldsymbol{Y})$
coincides with that of $i_{n}^{(P_{S})}(\overline{\boldsymbol{U}},\boldsymbol{Y}^{\prime})$.
Substituting $y\leftarrow y_{i}$ and $u\leftarrow u_{i}$ into \eqref{eq:DPC_Prop5}
and summing from $i=1$ to $n$, we see that $i_{n}^{(P_{S})}(\boldsymbol{u},\boldsymbol{y})$
depends on $(\boldsymbol{u},\boldsymbol{y})$ only through $\|\boldsymbol{u}\|^{2}$,
$\|\boldsymbol{y}\|^{2}$ and $\langle\boldsymbol{u},\boldsymbol{y}\rangle$.
Thus, since $\overline{\boldsymbol{U}}$ is circularly symmetric and
has a fixed magnitude, the distribution of $i_{n}^{(P_{S})}(\overline{\boldsymbol{U}},\boldsymbol{y})$
depends on $\boldsymbol{y}$ only through $\|\boldsymbol{y}\|^{2}$.
Writing $\boldsymbol{Y}=\boldsymbol{u}+(1-\alpha)\boldsymbol{s}+\boldsymbol{Z}$
and $\boldsymbol{Y}^{\prime}=\boldsymbol{X}^{\prime}+\boldsymbol{Z}$,
we see that conditioned on $(\boldsymbol{s},\boldsymbol{u})$, both
$\boldsymbol{Y}$ and $\boldsymbol{Y}^{\prime}$ are obtained by adding
the i.i.d. Gaussian vector $\boldsymbol{Z}$ to a vector whose power
is (almost surely or deterministically) equal to $\|\boldsymbol{u}+(1-\alpha)\boldsymbol{s}\|^{2}$.
Thus, the conditional distribution of $\|\boldsymbol{Y}\|^{2}$ coincides
with that of $\|\boldsymbol{Y}^{\prime}\|^{2}$, and we obtain \eqref{eq:DPC_SwapDistr}.

We now turn to the proof of \eqref{eq:DPC_Y'tail}--\eqref{eq:DPC_ChgMeasure}.
For the sake of notational brevity, we define $P_{Y}\triangleq P+P_{S}+1$,
and let $\mathcal{B}_{\epsilon}$ denote the set of sequences $\boldsymbol{y}^{\prime}$
such that $\big|\|\boldsymbol{y}^{\prime}\|^{2}-nP_{Y}\big|\le n\epsilon$.
By definition, $(\boldsymbol{Y}^{\prime}|\boldsymbol{s},\boldsymbol{u})$
is obtained by adding i.i.d. Gaussian noise to $(\boldsymbol{X}^{\prime}|\boldsymbol{s},\boldsymbol{u})$,
which in turn is uniform on a shell of power $n(P+P_{S})+\eta$ for
some $-\delta_{xs}\le\eta\le\delta_{xs}$ (see \eqref{eq:DPC_X'}
and \eqref{eq:DPC_Prop1}). Defining $f_{Y,n}^{(P_{S})}\sim N\big(0,P_{Y}+\frac{\eta}{n}\big)$,
Step 1 of the proof of \cite[Lemma 61]{Finite} states there exists
$\epsilon>0$ such that%
\footnote{More precisely, it was shown in \cite{Finite} that the ratio of the
densities of the \emph{norms} is upper bounded by a constant on $\mathcal{B}_{\epsilon}$.
Since we are considering circularly symmetric distributions, this
immediately implies that the same is true of the densities of the
sequences themselves.%
}
\begin{equation}
f_{\boldsymbol{Y}^{\prime}|\boldsymbol{S}\boldsymbol{U}}^{(P_{S})}(\boldsymbol{y}^{\prime}|\boldsymbol{s},\boldsymbol{u})\le K_{2}^{\prime}\prod_{i=1}^{n}f_{Y,n}^{(P_{S})}(y_{i}^{\prime})\label{eq:DPC_ChgMeasure1}
\end{equation}
for $\boldsymbol{y}^{\prime}\in\mathcal{B}_{\epsilon}$, where $K_{2}^{\prime}$
is a constant depending on $P_{Y}$ (see also \cite[Prop. 2]{MolavianJazi}).
As noted in \cite{Finite}, we can choose $\epsilon=1$ (or any $\epsilon\in(0,1)$).
The exponential decay in \eqref{eq:DPC_Y'tail} follows from the Chernoff
bound and the fact that $\mathbb{E}\big[\|\boldsymbol{Y}^{\prime}\|^{2}\,|\,\boldsymbol{s},\boldsymbol{u}\big]=nP_{Y}+\eta$
\cite[Eq. (417)]{Finite}.

To complete the proof of \eqref{eq:DPC_ChgMeasure}, we show that
for $\boldsymbol{y}^{\prime}\in\mathcal{B}_{\epsilon}$ we have $\prod_{i=1}^{n}f_{Y,n}^{(P_{S})}(y_{i}^{\prime})\le K_{2}^{\prime\prime}\prod_{i=1}^{n}f_{Y}^{(P_{S})}(y_{i}^{\prime})$
for some constant $K_{2}^{\prime\prime}$. We have from \eqref{eq:DPC_fY}
that $f_{Y}^{(P_{S})}\sim N(0,P_{Y})$, and hence
\begin{align}
\frac{\prod_{i=1}^{n}f_{Y,n}^{(P_{S})}(y_{i}^{\prime})}{\prod_{i=1}^{n}f_{Y}^{(P_{S})}(y_{i}^{\prime})} & =\sqrt{\frac{nP_{Y}}{nP_{Y}+\eta}}\exp\Bigg(-\frac{\|\boldsymbol{y}^{\prime}\|^{2}}{2}\bigg(\frac{1}{nP_{Y}+\eta}-\frac{1}{nP_{Y}}\bigg)\Bigg)\\
 & =\sqrt{\frac{1}{1+\frac{\eta}{nP_{Y}}}}\exp\Bigg(-\frac{\|\boldsymbol{y}^{\prime}\|^{2}}{2nP_{Y}}\bigg(\frac{1}{1+\frac{\eta}{nP_{Y}}}-1\bigg)\Bigg)\\
 & =\sqrt{\frac{1}{1+\frac{\eta}{nP_{Y}}}}\exp\Bigg(\frac{\|\boldsymbol{y}^{\prime}\|^{2}\eta}{2(nP_{Y})^{2}}\bigg(\frac{1}{1+\frac{\eta}{nP_{Y}}}\bigg)\Bigg)\label{eq:DPC_ChgMeasure4}
\end{align}
In the case that $\eta\in[-\delta_{xs},0)$, the desired result follows
since the argument to $\exp(\cdot)$ in \eqref{eq:DPC_ChgMeasure4}
is negative, and the subexponential prefactor tends to one. In the
case that $\eta\in[0,\delta_{xs}]$, the bound $\|\boldsymbol{y}^{\prime}\|^{2}\le n(P_{Y}+\epsilon)$
(within $\mathcal{B}_{\epsilon}$) implies that the argument to $\exp(\cdot)$
in \eqref{eq:DPC_ChgMeasure4} is upper bounded by $\frac{(P_{Y}+\epsilon)^{2}\delta_{xs}}{2P_{Y}^{2}}$,
which is a constant. It follows that \eqref{eq:DPC_ChgMeasure} holds
for all $\eta\in[-\delta_{xs},\delta_{xs}]$, as desired. 

Finally, the constants $\psi$ and $K_{2}$ in \eqref{eq:DPC_Y'tail}--\eqref{eq:DPC_ChgMeasure}
can be taken as independent of $P_{S}$ due to the fact that $P_{S}$
(and hence $P_{Y})$ is uniformly bounded within $\tilde{\mathcal{P}}_{n}$.

\subsubsection{Proof of Lemma \ref{lem:DPC_Property2}}

The evaluation of the moments of the information density is cumbersome
and similar to \cite[Appendix A]{PaperCMAC}, so we omit some of the
details.

We first consider the mean and variance. We write \eqref{eq:DPC_Prop5}
as
\begin{align}
i^{(P_{S})}(u,y) & =c_{0}+c_{1}(y+c_{2}u)^{2}+c_{3}y^{2}\\
 & =c_{0}+c_{1}c_{2}^{2}u^{2}+2c_{1}c_{2}uy+(c_{1}+c_{3})y^{2},\label{eq:DPC_in_alt}
\end{align}
where
\begin{align}
c_{0} & \triangleq\frac{1}{2}\log\frac{(P+P_{S}+1)(P+\alpha^{2}P_{S})}{PP_{S}(1-\alpha)^{2}+(P+\alpha^{2}P_{S})}\\
c_{1} & \triangleq-\frac{P+\alpha^{2}P_{S}}{2(PP_{S}(1-\alpha)^{2}+(P+\alpha^{2}P_{S}))}\\
c_{2} & \triangleq-\frac{P+\alpha P_{S}}{P+\alpha^{2}P_{S}}\\
c_{3} & \triangleq\frac{1}{2(P+P_{S}+1)}.
\end{align}
Substituting $y=u+(1-\alpha)s+z$ into \eqref{eq:DPC_in_alt}, we
obtain
\begin{equation}
i^{(P_{S})}(u,y)=c_{0}+d_{1}s^{2}+d_{2}u^{2}+d_{3}z^{2}+d_{4}su+d_{5}sz+d_{6}uz,\label{eq:DPC_in_alt2}
\end{equation}
where 
\begin{align}
d_{1} & \triangleq(c_{1}+c_{3})(1-\alpha)^{2}\\
d_{2} & \triangleq c_{1}c_{2}^{2}+2c_{1}c_{2}+(c_{1}+c_{3})\\
d_{3} & \triangleq c_{1}+c_{3}\\
d_{4} & \triangleq2c_{1}c_{2}(1-\alpha)+2(c_{1}+c_{3})(1-\alpha)\\
d_{5} & \triangleq2(c_{1}+c_{3})(1-\alpha)\\
d_{6} & \triangleq2c_{1}c_{2}+2(c_{1}+c_{3}).
\end{align}
Letting $Y=u+(1-\alpha)s+Z$ (which follows by combining $Y=x+s+Z$
and $x=u-\alpha s$) and taking the mean and variance of \eqref{eq:DPC_in_alt2}
with respect to $Z\sim N(0,1)$ for fixed $(s,u)$, we obtain
\begin{align}
\mathbb{E}\Big[i^{(P_{S})}(u,Y)\,|\, s,u\Big] & =c_{0}+d_{1}s^{2}+d_{2}u^{2}+d_{3}+d_{4}su\\
\mathrm{Var}\Big[i^{(P_{S})}(u,Y)\,|\, s,u\Big] & =\mathrm{Var}\Big[d_{3}Z^{2}+(d_{5}s+d_{6}u)Z\Big]\\
 & =2d_{3}^{2}+(d_{5}s+d_{6}u)^{2},
\end{align}
where we have used $\mathbb{E}[Z]=0$, $\mathrm{Var}[Z]=1$, $\mathrm{Var}[Z^{2}]=2$
and $\mathrm{Cov}[Z^{2},Z]=0$. It follows that
\begin{align}
\mathbb{E}\Big[i_{n}^{(P_{S})}(\boldsymbol{u},\boldsymbol{Y})\,|\,\boldsymbol{s},\boldsymbol{u}\Big] & =nc_{0}+d_{1}\|\boldsymbol{s}\|^{2}+d_{2}\|\boldsymbol{u}\|^{2}+nd_{3}+d_{4}\langle\boldsymbol{s},\boldsymbol{u}\rangle\label{eq:DPC_Efinal}\\
\mathrm{Var}\Big[i_{n}^{(P_{S})}(\boldsymbol{u},\boldsymbol{Y})\,|\,\boldsymbol{s},\boldsymbol{u}\Big] & =2nd_{3}^{2}+d_{5}^{2}\|\boldsymbol{s}\|^{2}+2d_{5}d_{6}\langle\boldsymbol{s},\boldsymbol{u}\rangle+d_{6}^{2}\|\boldsymbol{u}\|^{2}.\label{eq:DPC_Vfinal}
\end{align}
Using the definitions of the constants $c_{i}$ and $d_{i}$, it can
be verified from \eqref{eq:DPC_Efinal}--\eqref{eq:DPC_Vfinal} that,
for any $(\boldsymbol{s},\boldsymbol{u})$ such that $\|\boldsymbol{s}\|^{2}=nP_{S}$,
$\|\boldsymbol{u}\|^{2}=n(P+\alpha^{2}P_{S})$, and $\langle\boldsymbol{s},\boldsymbol{u}\rangle=n\alpha P_{S}$,
we have 
\begin{align}
\mathbb{E}\big[i_{n}^{(P_{S})}(\boldsymbol{u},\boldsymbol{Y})\,|\,\boldsymbol{s},\boldsymbol{u}\big] & =nI^{(P_{S})}(U;Y)\label{eq:DPC_Evec}\\
\mathrm{Var}\big[i_{n}^{(P_{S})}(\boldsymbol{u},\boldsymbol{Y})\,|\,\boldsymbol{s},\boldsymbol{u}\big] & =\frac{1}{2(1+P+P_{S})^{2}\big(PP_{S}(1-\alpha)^{2}+P+\alpha^{2}P_{S}\big)^{2}}\Bigg((P+\alpha P_{S})^{2}\bigg(\alpha^{2}P_{S}(2+P_{S})\nonumber \\
 & +P^{2}\Big(1+2(1-\alpha)^{2}P_{S}\Big)+2P\Big(1+\alpha P_{S}+P_{S}(1-\alpha)^{2}(2+P_{S})\Big)\bigg)\Bigg)\\
 & \triangleq nV(P_{S}).\label{eq:DPC_Vvec2}
\end{align}
The distribution of $(\boldsymbol{S},\boldsymbol{U})$ under consideration
does not ensure that the equalities $\|\boldsymbol{u}\|^{2}=n(P+\alpha^{2}P_{S})$
and $\langle\boldsymbol{s},\boldsymbol{u}\rangle=n\alpha P_{S}$ hold.
However, they do hold to within an additive $O(1)$ term; see \eqref{eq:DPC_SetTn}
and \eqref{eq:DPC_Prop3}. Since the right-hand sides of \eqref{eq:DPC_Efinal}--\eqref{eq:DPC_Vfinal}
are linear in $\|\boldsymbol{s}\|^{2}$ and $\langle\boldsymbol{s},\boldsymbol{u}\rangle$,
we conclude that \eqref{eq:DPC_Evec} and \eqref{eq:DPC_Vvec2} hold
for all $(\boldsymbol{s},\boldsymbol{u})$ on the support of $f_{\boldsymbol{S}\boldsymbol{U}}^{(P_{S})}$
upon adding $O(1)$ to the right-hand sides. A direct substitution
of $\alpha=\frac{P}{1+P}$ reveals that $V(P_{S})=\frac{P(2+P)}{2(1+P)^{2}}$
for all $P_{S}$, and we have thus proved \eqref{eq:DPC_CondAvg}--\eqref{eq:DPC_CondVar}.

It remains to prove \eqref{eq:DPC_CondThird}. To this end, we use
\eqref{eq:DPC_in_alt2} to write $i_{n}^{(P_{S})}(\boldsymbol{u},\boldsymbol{Y})$
(given $(\boldsymbol{s},\boldsymbol{u})$) as
\begin{equation}
i_{n}^{(P_{S})}(u,y)=nd_{0}+d_{1}\|\boldsymbol{s}\|^{2}+d_{2}\|\boldsymbol{u}\|^{2}+d_{3}\|\boldsymbol{Z}\|^{2}+d_{4}\langle\boldsymbol{s},\boldsymbol{u}\rangle+d_{5}\langle\boldsymbol{s},\boldsymbol{Z}\rangle+d_{6}\langle\boldsymbol{u},\boldsymbol{Z}\rangle.
\end{equation}
Since $\boldsymbol{Z}$ is i.i.d. Gaussian, the distributions of the
last two terms only depend on $\|\boldsymbol{s}\|^{2}$ and $\|\boldsymbol{u}\|^{2}$.
Thus, the statistics of $i_{n}^{(P_{S})}(\boldsymbol{U},\boldsymbol{Y})$
given $(\boldsymbol{S},\boldsymbol{U})=(\boldsymbol{s},\boldsymbol{u})$
depends on $(\boldsymbol{s},\boldsymbol{u})$ only through $\|\boldsymbol{u}\|^{2}$,
$\|\boldsymbol{s}\|^{2}$ and $\langle\boldsymbol{s},\boldsymbol{u}\rangle$.
We thus obtain \eqref{eq:DPC_CondThird} in the same way as \cite[Appendix A]{PaperCMAC}
by choosing $(\boldsymbol{s}^{\prime},\boldsymbol{u}^{\prime})$ to
attain the same powers and correlation as $(\boldsymbol{s},\boldsymbol{u})$
(thus yielding the same statistics of $i_{n}^{(P_{S})}$), while having
entries which are uniformly bounded for all $n$.

\section*{Acknowledgments}

I would like to thank Vincent Tan for many helpful suggestions which
greatly helped in making the proofs more rigorous and complete. In
particular, Corollary \ref{cor:GP_DecStateInfo} was pointed out by
him.

\bibliographystyle{IEEEtran}
\bibliography{12-Paper,18-MultiUser,18-SingleUser,35-Other}

\end{document}